\documentclass[10pt,conference,letterpaper]{IEEEtran}
\input{psfig.sty}
\input{epsf.sty}
 \usepackage{hyperref} 
\usepackage{fancybox}   
\usepackage{psfig}      
\usepackage{graphicx}
\usepackage{amsmath}
\usepackage{color}
\usepackage{epsfig}
\usepackage{amssymb}
\usepackage{verbatim}
\usepackage{enumitem}
\usepackage{bbm}
\usepackage{lipsum}

\IEEEoverridecommandlockouts

\newcommand{\remove}[1]{}

\newtheorem{theorem}{Theorem}

\newtheorem{lemma}{Lemma}

\newtheorem{remark}{Remark}

\newtheorem{algorithm}{Algorithm}
\newtheorem{assumption}{Assumption}

\usepackage{setspace}
\usepackage{subfigure}

\begin{document}

 \title{Optimal Dynamic Sensor Subset Selection for Tracking a  Time-Varying Stochastic Process
 \thanks{Parts of this paper have been published in previous conferences; see \cite{arpan-globecom2017-techreport}, \cite{inform-techreport}.}
 \thanks{Arpan Chattopadhyay and  Urbashi Mitra are with Ming Hsieh Department of Electrical Engineering, University of 
 Southern California, Los Angeles, USA. Email: \{achattop,ubli\}@usc.edu}\\
 \thanks{This work was funded by the following grants: ONR N00014-15-1-2550, NSF CNS-1213128, NSF CCF-1410009, AFOSR
FA9550-12-1-0215, and NSF CPS-1446901.}
}

\author{
Arpan~Chattopadhyay, Urbashi~Mitra \\
}

\maketitle
\thispagestyle{empty}

\begin{abstract}
Motivated by the Internet-of-things and sensor networks for cyberphysical systems, the problem of dynamic sensor activation for the tracking of a time-varying process is examined. The tradeoff is between energy efficiency, which decreases with the number of active sensors, and fidelity, which increases with the number of active sensors.  The problem of minimizing the time-averaged mean-squared error over infinite horizon is examined under the constraint of  the mean number of active sensors.  The proposed methods artfully combine three key ingredients: Gibbs sampling, stochastic approximation for learning, and modifications to consensus algorithms to create a high performance, energy efficient tracking mechanisms with active sensor selection.  The following progression of scenarios are considered: centralized tracking of an i.i.d. process; distributed tracking of an i.i.d. process and finally distributed tracking of a Markov chain.  The challenge of the i.i.d. case is that the process has a distribution parameterized by a known or  unknown parameter which must be learned.  The key theoretical results  prove that the proposed algorithms converge to local optima for the two i.i.d process cases; numerical results suggest that global optimality is in fact achieved. The proposed distributed tracking algorithm for a Markov chain, based on Kalman-consensus filtering and stochastic approximation,  is seen to offer an error performance comparable to that of a competetive centralized Kalman filter.
\end{abstract}

\section{Introduction}\label{section:introduction}
Controlling and monitoring physical processes via sensed data are integral parts of internet-of-things (IOT), cyber-physical systems, and defense  with applications to industrial process monitoring and control, localization, tracking of mobile objects,  environmental monitoring, system identification and disaster management.  In such applications, sensors are simultaneously resource constrained (power and/or bandwdith) and tasked to achieve high performance sensing, control, communication, and tracking.  Wireless sensor networks must further contend with interference and fading.  One strategy for balancing resource use with performance is to activate a subset of the total possible number of sensors to limit both computation as well as bandwidth use.
  
Herein, we address the fundamental problem of optimal dynamic sensor subset selection for tracking a time-varying stochastic process.   We first examine the centralized tracking of an iid process with an unknown, parametric distribution which serves as a benchmark for the first extension to decentralized tracking of this process.  For both  proposed algorithms, almost sure convergence to local optima can be proven.  Next a distributed algorithm for tracking a Markov process with a known probability transition matrix is developed.  All algorithms are numerically validated.

\subsection{Related Literature}
\label{subsection:related-literature}
Optimal sensor subset selection problems can be broadly classified into two 
categories: (i) optimal sensor subset selection for static data with known
prior distribution, but unknown realization, and (ii)  dynamic sensor subset
selection  to track a time-varying stochastic process. There have been several
recent attempts to solve the first problem; see \cite{wang-etal16efficient-observation-selection, arpan-globecom2017-techreport}.  
This problem poses two major challenges: (i) computing the estimation
error given the observations from a subset of sensors, and (ii) finding the
optimal sensor subset from exponentially many number of subsets. In \cite{wang-etal16efficient-observation-selection}, a tractable lower bound on performance addressed the first challenge and a greedy algorithm addressed the second.  In \cite{arpan-globecom2017-techreport} only the second challenge is addressed via  Gibbs sampling  approach. 

Dynamic sensor subset selection  for  a time-varying stochastic process is considered in {\em e.g.} \cite{daphney-etal14active-classification-pomdp, daphney-etal13energy-efficient-sensor-selection, krishnamurthy07structured-threshold-policies, wu-arapostathis08optimal-sensor-querying, gupta-etal06stochastic-sensor-selection-algorithm, bertrand-moonen10sensor-selection-linear-mmse}.  Markov process tracking with centralized controllers for single (\cite{krishnamurthy07structured-threshold-policies}) or multiple sensor selection with energy constraints  and sequential decision making (\cite{daphney-etal14active-classification-pomdp, daphney-etal13energy-efficient-sensor-selection}) have been previously studied. The optimal policy and its structural properties for a special case of dynamic sensor selection over an infinite horizon was examined in \cite{wu-arapostathis08optimal-sensor-querying}. Single sensor selection with the broadcast of sensor data  is studied in  \cite{gupta-etal06stochastic-sensor-selection-algorithm}.  To our knowledge, the combination of Gibbs sampling ({\em e.g.} \cite{breamud99gibbs-sampling})  and stochastic approximation ({\em e.g.} \cite{borkar08stochastic-approximation-book}) has not been previously applied to iid process tracking as we do herein. The paper \cite{schnitzler-etal15sensor-selection-crowdsensing}, using Thompson sampling, has solved the problem of {\em centralized} tracking of a linear Gaussian process (with unknown noise statistics) via active sensing.

Given our consideration of dynamic sensor subset selection for {\em distributed} tracking of a Markov process, traditional Kalman filtering is not applicable; however, there is much prior, recent art on the developmemt of consensus-based Kalman filtering \cite{das-moura17consensus-innovation-distributed-KF,  kar-moura11gossip-and-DKF,   saber09kalman-consensus-optimality-stability,   xiao-etal05scheme-sensor-fusion-average-consensus, tzoumas-etal16sensor-scheduling-batch-state-estimation}  which we further adapt to work with our Gibbs sampling/stochastic approximation approach for the iid case.  This distributed tracking problem does not appear to be heavily studied.  In \cite{gupta-etal06stochastic-sensor-selection-algorithm} perfect information sharing between sensors is assumed, which is impractical and \cite{michelusi} assumes that the estimation is done in a centralized fashion, while sensors make decentralized decisions about whether to sense or communicate to the fusion center.  In \cite{michelusi}, the sparsity of the Markov process is also exploited.

\subsection{Our Contribution}\label{subsection:organization}
In this paper, we make the following contributions:

\begin{enumerate}
\item    A centralized tracking and learning algorithm for an iid process   with an unknown, but parametric distribution is developed.  In particular, Gibbs sampling minimizes computational complexity with stochastic approximation to achieve the mean number of activated sensors constraint.  Furthermore, simultaneous perturbation stochastic approximation (SPSA) is employed for parameter estimation obviating the need for the expectation-maximization algorithm.  A challenge we overcome in the analysis,  is handling updates at different time scales. As a precursor to the algorithm for unknown parameters, algorithms have been developed for a simpler version of the problem when the parameter of the distribution is known.

\item  The centralized algorithm, which serves as a benchmark, is adapted to the distributed case by exploiting partial consensus. Our partial consensus is novel in that  SPSA is again employed to learn the optimal consensus gains adaptively.

\item   A trick for ensuring that all sensors employ similar sampling strategies is to have each sensor use the same seed in a random number generator.

\item  For both the centralized and distributed algorithms, we can prove almost sure convergence to local optima.  Furthermore, we an prove that the resources needed for communication and learning can be made arbitrarily small, by exploiting properties of the multi-scale updates.

\item  Our final generalization to  is to develop an algorithm for sensor subset selection for the decentralized tracking of a Markov chain with known probability transition matrix.  We adapt methods for Kalman consensus filtering to our framework with Gibbs sampling and stochastic approximation.

\item  Numerical results show that the decentralized scheme for a Markov chain performs close to that of the centralized scheme. Numerical results also show a tradeoff between performance and message exchange for learning.  Furthermore, the numerical results show that global (not local) optima are achieved in tracking iid process.
\end{enumerate}

\subsection{Organization}
The rest of the paper is organized as follows. The system model  is described in Section~\ref{section:system-model}. Centralized tracking of an iid process with known distribution is described in Section~\ref{section:gibbs-sampling-unconstrained-problem}.  Section~\ref{section:iid-data} deals with centralized tracking of an iid process having a parametric distribution with unknown parameters. Distributed tracking of iid process is discussed in Section~\ref{section:iid-data-distributed-estimation}. Distributed tracking of a Markov chain is described in Section~\ref{section:markovian-data}. Numerical results are presented in Section~\ref{section:numerical-results}, followed by the conclusion in Section~\ref{section:conclusion}.

\section{System Model}\label{section:system-model}

\subsection{Network, data and sensing model}\label{subsection:network-data-model}
We consider a  connected single or multi-hop wireless sensor network. The sensor nodes are denoted by the set   $\mathcal{N}=\{1,2,\cdots,N\}$.  It might also be possible that the connectivity in the network is maintained via a few relay nodes, but we ignore this possibility for the ease of analysis. There can be a fusion center (connected to all sensors via single hop wireless links) responsible for all control or estimation operations in the network, or, alternatively, the sensors  can operate autonomously in a multihop mesh network.

The physical process under measurement is denoted by  $\{X(t)\}_{t \geq 0}$, where $t$ is a discrete time index and $X(t) \in \mathbb{R}^{q \times 1}$. We consider two models for the evolution of $\{X(t)\}_{t \geq 0}$:

\begin{itemize}
\item {\em IID model:}  $\{X(t)\}_{t \geq 0}$ is an i.i.d. process with a parametric distribution $p_{\theta_0}(X)$, where the {\em unknown} parameter $\theta_0$   needs to be be learnt via the  measurements. {\em $\theta_0$ lies inside the interior of a compact subset $\Theta \subset \mathbb{R}^d$.}
\item {\em Markov model:}  $\{X(t)\}_{t \geq 0}$ is a finite-state   ergodic Markov chain with known transition probability matrix.
\end{itemize}
 At time $t$, if a sensor~$k$ is used to sense the process, then the observation at sensor~$k$ is provided by a $r$-dimensional column vector  $z_k(t)=H_k X(t)+v_k(t)$ where $H_k $ is an observation matrix of appropriate dimension, and $v_k(t)$ is a Gaussian vector;  the observation noise $v_k(t)$ is assumed to be independent across $k$ and i.i.d. across $t$. 
 
 Let $B(t) \in \{0,1\}^{1 \times N}$ be a vector where $B_k(t)$ is the indicator that the sensor~$k$ is activated at time~$t$; $B_k(t)=1$ if sensor~$k$ is active at time~$t$, or else $B_k(t)=0$. The decision to activate any sensor for sensing and communicating the observation    is taken either by the fusion center or by the sensor itself in the absence of a fusion center. We denote by $\mathcal{B}:=\{0,1\}^N$ the set of all possible configurations in the network, and by $B$ a generic configuration. Clearly, $B(t) \in \mathcal{B}$. Each configuration represents a set of activated sensors.  $B_{-j} \in \{0,1\}^{N-1}$ is used to represent the configuration $B$ with its $j$-th entry removed.

The observation made by sensor~$k$ at time $t$ is  $Z_k(t)=B_k(t) z_k(t)$. We define $Z(t):=\{Z_k(t): 1 \leq k \leq N\}\}$.

\subsection{Centralized estimation problem}\label{subsection:centralized-estimation-problem}
The estimate of $X(t)$ at the fusion center (connected to all sensors via direct wireless links) is denoted by $\hat{X}(t)$. We denote by $\mathcal{H}_p(t)$ the information available at time $t$ at the fusion center about the history of observations, activations and estimates up to time $(t-1)$, before the fusion center determines $B(t)$. On the other hand, we define $\mathcal{H}_c(t):=\{B(t); Z(t); \theta(t)\}$ where $\theta(t)$ is the current estimate of $\theta_0$; $\mathcal{H}_c(t)$ denotes the information used by the fusion center at time $t$ to estimate $\hat{X}(t)$.  For i.i.d. time varying process, $\mathcal{H}_c(t)$ is sufficient to estimate $X(t)$ and obtain $\hat{X}(t)$, and $\mathcal{H}_c(t)$ will be available only after deciding the activation vector $B(t)$ and collecting all observations.  In order to optimally decide $B(t)$, the fusion center needs  knowledge about the performance of all configurations in the past. Hence, $\mathcal{H}_p(t)$ and $\mathcal{H}_c(t)$ have two different information structures. However, we will see that, our Gibbs sampling algorithm determines $B(t)$ by using only a sufficient statistic   calculated iteratively in each slot.

 The information structure $\mathcal{H}_c(t)$ used to track a Markov chain will be different,  which we will see  in Section~\ref{section:markovian-data}.

We define a policy $\mu=\{(\mu_1,\mu_2\}$ as a pair of mappings, where $\mu_1(\mathcal{H}_p(t))=B(t)$ and $\mu_2(\mathcal{H}_c(t))=\hat{X}(t)$. Our first goal is to solve the following centralized   problem of minimizing the time-average mean squared error (MSE) subject to a constraint on the mean number of sensors active per unit time:

\footnotesize
\begin{align}
&& \mu^*=\arg \min_{\mu} \limsup_{t \rightarrow \infty} \frac{1}{t} \sum_{\tau=1}^t \mathbb{E}_{\mu} ||X(\tau)-\hat{X}(\tau)||^2  \nonumber\\
&& s.t.\,\,\,\, \limsup_{t \rightarrow \infty} \frac{1}{t} \sum_{\tau=1}^t  \mathbb{E}_{\mu} ||B(\tau)||_1 \leq \bar{N} \tag{P1} \label{eqn:centralized-constrained-problem}
\end{align}
\normalsize
\subsection{Distributed estimation problem}\label{subsection:distributed-estimation-problem}
In the absence of a fusion center in a multi-hop network, the estimate of $X(t)$ at sensor~$k$ is denoted by $\hat{X}^{(k)}(t)$. We denote by $\mathcal{H}_p^{(k)}(t)$ the information available at time $t$ at sensor~$k$ about the history before the sensor determines   $B_k(t)$, and by $\mathcal{H}_c^{(k)}(t)$ the information available at sensor~$k$ at time $t$ just before $\hat{X}^{(k)}(t)$ is estimated. 

We define a policy $\mu=\{(\mu_1^{(k)},\mu_2^{(k)})\}_{1 \leq k \leq N}$, where $\mu_1^{(k)}(\mathcal{H}_p^{(k)}(t))=B_k(t)$ and $\mu_2^{(k)}(\mathcal{H}_c^{(k)}(t))=\hat{X}^{(k)}(t)$.

We seek to solve the following distributed  problem:

\footnotesize
\begin{align}
 \mu^*&=&\arg \min_{\mu}  \limsup_{t \rightarrow \infty} \frac{1}{t} \sum_{\tau=1}^t  \frac{1}{N} \sum_{k=1}^N \mathbb{E}_{\mu} ||X(\tau)-\hat{X}^{(k)}(\tau)||^2  \nonumber\\
&&  s.t.\,\,\,\,   \limsup_{t \rightarrow \infty} \frac{1}{t} \sum_{\tau=1}^t  \mathbb{E}_{\mu} ||B(\tau)||_1 \leq \bar{N} \tag{P2}
\label{eqn:distributed-constrained-problem}
\end{align}
\normalsize

\section{Centralized Tracking of IID proces: known $\theta_0$}
\label{section:gibbs-sampling-unconstrained-problem}
In this section, we  provide an algorithm for solving the centralized problem \eqref{eqn:centralized-constrained-problem} when $X(t) \sim p_{\theta_0}(\underline{x)}$ i.i.d. This is done by relaxing \eqref{eqn:centralized-constrained-problem} by a Lagrange multiplier. Though our final goal is to track an i.i.d. process with unknown $\theta_0$, we discuss algorithms for known $\theta_0$ in this section, as a precursor to the algorithms developed in subsequent sections for unknown $\theta_0$ and also tracking a Markov chain. Also, extension to distributed tracking will be discussed in Section~\ref{section:iid-data-distributed-estimation} for unknown $\theta_0$, and hence will be omitted in this section.

\subsection{The relaxed version of the constrained problem}\label{subsection:relaxed-version-centralized-constrained-problem-iid-data}
We relax \eqref{eqn:centralized-constrained-problem} by using a Lagrance multiplier $\lambda$ and obtain the following unconstrained problem:

\begin{equation} \label{eqn:centralized-unconstrained-problem} 
\mu^*=\arg \min_{\mu} \limsup_{t \rightarrow \infty} \frac{1}{t} \sum_{\tau=1}^t \mathbb{E}_{\mu} \bigg(  ||X(\tau)-\hat{X}(\tau)||^2 + \lambda  ||B(\tau)||_1 \bigg) \tag{P3}
\end{equation}
\normalsize
The multiplier $\lambda \geq 0$ can be viewed as the cost incurred for activating a sensor at any time instant. 
Now, since \eqref{eqn:centralized-unconstrained-problem} is an unconstrained problem and $X(t)$ is i.i.d. across $t$, there exists one optimizer $B^* \in \mathcal{B}$ (not necessarily unique) for the problem \eqref{eqn:centralized-unconstrained-problem}; if the configuration $B^*$ is chosen at each $t$, the minimum cost of \eqref{eqn:centralized-unconstrained-problem} can be achieved by the law of large numbers. Hence, for known $\theta_0$, the problem \eqref{eqn:centralized-unconstrained-problem} can be equivalently written as:
\begin{equation}\label{eqn:centralized-unconstrained-problem-with-f-and-h} 
\arg \min_{B \in \mathcal{B}} \underbrace{ \underbrace{\mathbb{E}_{\mu_2,B} ||X(\tau)-\hat{X}(\tau)||^2}_{:=f(B)}+ \lambda ||B||_1 }_{:=h(B)} \tag{P4}
\end{equation}
The following   result tells us how to choose the optimal $\lambda^*$ to solve   \eqref{eqn:centralized-constrained-problem}. 

\noindent\fbox{
    \parbox{0.46 \textwidth}{
\begin{theorem}\label{theorem:relation-between-constrained-and-unconstrained-problems}
Consider problem \eqref{eqn:centralized-constrained-problem} and  its relaxed version \eqref{eqn:centralized-unconstrained-problem-with-f-and-h}. If there exists a Lagrange multiplier $\lambda^* \geq 0$ and a $B^* \in \mathcal{B}$, such that an optimal configuration for \eqref{eqn:centralized-unconstrained-problem-with-f-and-h}  under  $\lambda=\lambda^*$ is $B^*$, and the constraint in \eqref{eqn:centralized-constrained-problem}  is satisfied with equality under the pair 
$(B^*,\lambda^*)$, then $B^*$ is an optimal configuration for \eqref{eqn:centralized-constrained-problem}.

In case there exist multiple configurations $B_1^*,B_2^*, \cdots, B_m^*$, a multiplier  $\lambda^* \geq 0$, and a probability mass function $(p_1,p_2,\cdots,p_m)$  such that (i) each of $B_1^*,B_2^*,\cdots, B_m^*$ is optimal for problem~\eqref{eqn:centralized-unconstrained-problem-with-f-and-h}  under $\lambda^*$, and (ii) $\sum_{i=1}^m p_i ||B_i^*||_1=\bar{N}$,  then an optimal solution for \eqref{eqn:centralized-constrained-problem} is to choose one configuration from $B_1^*,B_2^*,\cdots, B_m^*$ with probability mass function  $(p_1,p_2,\cdots,p_m)$.
\end{theorem}
}}
\begin{proof}
See Appendix~\ref{appendix:proof-of-relation-between-constrained-and-unconstrained-problems}.
\end{proof}
\begin{remark}
Theorem~\ref{theorem:relation-between-constrained-and-unconstrained-problems} allows us to obtain a solution for \eqref{eqn:centralized-constrained-problem} from the solution of   by choosing an appropriate $\lambda^*$; this will be elaborated upon in  Section~\ref{subsection:gibbs-stochastic-approximation-unconstrained-problem}.
\end{remark}

\subsection{Basics of Gibbs sampling for known $p_{\theta_0}(\cdot)$}\label{subsection:finite-beta-gibbs-sampling-unconstrained-problem}
Finding the optimal solution of  \eqref{eqn:centralized-unconstrained-problem-with-f-and-h} requires us to search over  $2^N$ possible configurations and to compute MMSE for each of these configurations.  Hence, we propose  Gibbs sampling based algorithms to avoid this $O(2^N)$ computation.

Let us define the probability distribution   $\pi_{\beta}(\cdot)$ over $\mathcal{B}$ as (with $\beta>0$):
\begin{equation}\label{eqn:definition-of-Gibbs-distribution}
\pi_{\beta}(B):=\frac{e^{-\beta h(B)}}{\sum_{B' \in \mathcal{B}}e^{-\beta h(B')}}:=\frac{e^{-\beta h(B)}}{Z_{\beta}}.
\end{equation}

Following the terminology in statistical physics,  we call   $\beta$ the {\em inverse temperature}, and $Z_{\beta}$ the {\em partition function}. $h(B)$ is viewed as the {\em energy under configuration $B$. } 
Now, $\lim_{\beta \uparrow \infty} \sum_{ B \in \arg \min_{A \in \mathcal{B}} h(A) } \pi_{\beta}(B)=1$. Hence, if a configuration $B(t)$ is selected at each  time  $t$ with probability distribution $\pi_{\beta}(\cdot)$ for sufficiently large $\beta>0$, then $B(t)$ will belong to the set of minimizers of \eqref{eqn:centralized-unconstrained-problem} with high probability. However, computing   $Z_{\beta}$ requires $2^N$ addition operations; hence, we use a sequential subset selection algorithm based on Gibbs sampling (see  \cite[Chapter~$7$]{breamud99gibbs-sampling})  in order to avoid explicit computation of $Z_{\beta}$ while picking $X(t) \sim p_{\theta_0}(\cdot)$. 
\noindent\fbox{
    \parbox{0.46 \textwidth}{
\begin{algorithm}
Start with an initial configuration $B(0)$. At time~$t$, pick a random sensor~$j_t$ uniformly from the set of all sensors.    Choose $B_{j_t}(t)=1$ with probability 
$p(t):=\frac{ e^{-\beta h(B_{-j_t}(t-1),1)}}{e^{-\beta h(B_{-j_t}(t-1),1)}+e^{-\beta h(B_{-j_t}(t-1),0)}}$ and choose $B_{j_t}(t)=0$ with probability 
$(1-p(t))$. For $k  \neq j_t$,  choose $B_k(t)=B_k(t-1)$. Activate the sensors according to $B(t)$.
\label{algorithm:basic-gibbs-for-known-distribution-iid-data}
\end{algorithm}
}}

\noindent\fbox{
    \parbox{0.46 \textwidth}{
\begin{theorem}\label{theorem:convergence-basic-Gibbs-sampling}
Under Algorithm~\ref{algorithm:basic-gibbs-for-known-distribution-iid-data},  $\{B(t)\}_{t \geq 0}$ is a reversible, ergodic, time-homogeneous Markov chain  with stationary distribution $\pi_{\beta}(\cdot)$.
\end{theorem}
}}
\begin{proof}
Follows from the theory  in \cite[Chapter~$7$]{breamud99gibbs-sampling}).
\end{proof}
\begin{remark}
Theorem~\ref{theorem:convergence-basic-Gibbs-sampling} tells us that if the fusion center runs  Algorithm~\ref{algorithm:basic-gibbs-for-known-distribution-iid-data} and reaches the steady state distribution of the Markov chain $\{B(t)\}_{t \geq 0}$, then the configuration chosen by the algorithm will have distribution $\pi_{\beta}(\cdot)$. For very large  $\beta>0$, if one runs $\{B(t)\}_{t \geq 0}$ for a sufficiently long, finite time $T$, then the terminal state $B_{T}$ will belong to 
$\arg \min_{B \in \mathcal{B}} h(B)$ with high probability. Also, by the ergodicity of $\{B(t)\}_{t \geq 0}$, the time-average occurence rates of all configurations match the distribution $\pi_{\beta}(\cdot)$ almost surely.
\end{remark}

\subsection{The exact solution}\label{subsection:growing-beta-gibbs-sampling-unconstrained-problem}
Algorithm~\ref{algorithm:basic-gibbs-for-known-distribution-iid-data}  is operated with a fixed $\beta$; but  the optimal soultion of the unconstrained problem~\eqref{eqn:centralized-unconstrained-problem} can only be  obtained with 
$\beta \uparrow \infty$; this is done by updating  $\beta$  at a slower time-scale than the iterates of Algorithm~\ref{algorithm:basic-gibbs-for-known-distribution-iid-data}. 

\noindent\fbox{
    \parbox{0.46 \textwidth}{
\begin{algorithm}
This algorithm is same as Algorithm~\ref{algorithm:basic-gibbs-for-known-distribution-iid-data} except that at   time $t$, we use  
$\beta(t):=\beta(0) \log (1+t)$   to compute the update probabilities, where $\beta(0)>0$, $\beta(0) N \Delta<1$, and   $\Delta:=\max_{B \in \mathcal{B}, A \in \mathcal{B}}|h(B)-h(A)|$.
\label{algorthm:gibbs-sampling-with-increasing-beta}
\vspace{2mm}
\end{algorithm}
}}

\noindent\fbox{
    \parbox{0.46 \textwidth}{
\begin{theorem}\label{theorem:result-on-weak-and-strong-ergodicity}
Under Algorithm~\ref{algorthm:gibbs-sampling-with-increasing-beta}, the Markov chain $\{B(t)\}_{t \geq 0}$ is strongly ergodic, and the limiting probability distribution satisfies $\lim_{t \rightarrow \infty} \sum_{A \in \arg \min_{C \in \mathcal{B}} h(C) }\mathbb{P}(B(t)=A)=1$.
\end{theorem}
}}
\begin{proof}
See Appendix~\ref{appendix:proof-of-weak-and-strong-ergodicity}.  
We have used the notion of weak and strong ergodicity of time-inhomogeneous Markov chains from 
\cite[Chapter~$6$, Section~$8$]{breamud99gibbs-sampling}), which is  provided in Appendix~\ref{appendix:weak-and-strong-ergodicity}. The proof is similar to the proof of one theorem in  \cite{chattopadhyay-etal16gibbsian-caching-arxiv}, but is given here for completeness.
\end{proof}
\begin{remark}
Theorem~\ref{theorem:result-on-weak-and-strong-ergodicity} shows that we can solve \eqref{eqn:centralized-unconstrained-problem} {\em exactly} if we run Algorithm~\ref{algorthm:gibbs-sampling-with-increasing-beta} for infinite time, in contrast with Algorithm~\ref{algorithm:basic-gibbs-for-known-distribution-iid-data} which   provides an approximate solution.
\end{remark}
\begin{remark}
For  i.i.d. time varying $\{ X(t) \}_{t \geq 0}$ with known joint distribution, we can either: (i) find the optimal configuration $B^*$ using Algorithm~\ref{algorthm:gibbs-sampling-with-increasing-beta}   and use $B^*$ for ever, or (ii) run Algorithm~\ref{algorthm:gibbs-sampling-with-increasing-beta}  at the same timescale as $t$, and use the running configuration $B(t)$ for sensor activation; both schemes will minimize the  cost in \eqref{eqn:centralized-unconstrained-problem}. By the strong ergodicity of $\{B(t)\}_{t \geq 0}$, optimal cost will be achieved for \eqref{eqn:centralized-unconstrained-problem} under Algorithm~\ref{algorthm:gibbs-sampling-with-increasing-beta}.
\end{remark}

\subsubsection{Convergence rate of Algorithm~\ref{algorithm:basic-gibbs-for-known-distribution-iid-data}}
Let $\pi^{(t)}$ denote the probability distribution of $B(t)$ under Algorithm~\ref{algorithm:basic-gibbs-for-known-distribution-iid-data}.  
Let us consider the transition probability matrix $P$ of the Markov chain $\{Y(l)\}_{l \geq 0}$ with  $Y(l)=B(lN)$, under Algorithm~\ref{algorithm:basic-gibbs-for-known-distribution-iid-data}. Let us recall the definition of the Dobrushin's ergodic coefficient $\delta(P)$ from \cite[Chapter~$6$, Section~$7$]{breamud99gibbs-sampling} for the matrix $P$; using a method similar to that of the proof of Theorem~\ref{theorem:result-on-weak-and-strong-ergodicity}, we can show that $\delta(P) \leq ( 1-\frac{ e^{-\beta N \Delta} }{N^N})$. 
  Then, by \cite[Chapter~$6$, Theorem~$7.2$]{breamud99gibbs-sampling}, we can say that under  Algorithm~\ref{algorithm:basic-gibbs-for-known-distribution-iid-data}, we have $d_V(\pi^{(lN)},\pi_{\beta}) \leq d_V(\pi^{(0)},\pi_{\beta}) \bigg( 1-\frac{ e^{-\beta N \Delta} }{N^N}  \bigg)^l$. We can prove similar bounds for any $t=lN+k$, where $0 \leq k \leq N-1$.
  
  Unfortunately, we are not aware of such a closed-form bound for Algorithm~\ref{algorthm:gibbs-sampling-with-increasing-beta}.

\begin{remark}
Clearly, under Algorithm~\ref{algorithm:basic-gibbs-for-known-distribution-iid-data}, the convergence rate decreases as $\beta$ increases. Hence, there is a trade-off between convergence rate and accuracy of the solution in this case. Also, the rate of convergence decreases with $N$. For Algorithm~\ref{algorthm:gibbs-sampling-with-increasing-beta}, the convergence rate is expected to decrease with time. 
\end{remark}

\subsection{Gibbs sampling and stochastic approximation based approach to solve the constrained problem}
\label{subsection:gibbs-stochastic-approximation-unconstrained-problem}

In Section~\ref{subsection:finite-beta-gibbs-sampling-unconstrained-problem} and Section~\ref{subsection:growing-beta-gibbs-sampling-unconstrained-problem}, we presented Gibbs sampling based algorithms for  \eqref{eqn:centralized-unconstrained-problem}. Now we provide an algorithm that updates $\lambda$ with time in order to meet the constraint in 
\eqref{eqn:centralized-constrained-problem} with equality, and thereby solves \eqref{eqn:centralized-constrained-problem}  (via Theorem~\ref{theorem:relation-between-constrained-and-unconstrained-problems}). 

\noindent\fbox{
    \parbox{0.46 \textwidth}{
\begin{lemma}\label{lemma:active-sensors-decreasing-in-lambda}
For the unconstrained problem~\eqref{eqn:centralized-unconstrained-problem-with-f-and-h}, the optimal mean number of active sensors, $\mathbb{E}_{\mu_2}||B^*||_1$, decreases with $\lambda$. Similarly, the optimal 
error, $\mathbb{E}_{\mu_2}f(B^*)$, increases with $\lambda$.
\end{lemma}
}}
\begin{proof}
See Appendix~\ref{appendix:proof-of-active-sensors-decreasing-in-lambda}.
\end{proof}
\begin{remark}
The optimal mean number of active sensors, $\mathbb{E}_{\mu_2}||B^*||_1$, for the unconstrained problem~\eqref{eqn:centralized-unconstrained-problem-with-f-and-h} is a decreasing staircase function of $\lambda$, where each point of discontinuity is associated with a change in the optimizer $B^*(\lambda)$.
\end{remark}
Lemma~\ref{lemma:active-sensors-decreasing-in-lambda} provides an intuition about how to update $\lambda$ in  Algorithm~\ref{algorithm:basic-gibbs-for-known-distribution-iid-data} or in Algorithm~\ref{algorthm:gibbs-sampling-with-increasing-beta} in order  to solve \eqref{eqn:centralized-constrained-problem}.   We  seek to provide one  algorithm which updates $\lambda(t)$ at each time instant, based on the number of active sensors in the previous time instant. In order to maintain the necessary timescale difference between the $\{B(t)\}_{t \geq 0}$ process and the $\lambda(t)$ update process, we use stochastic approximation (\cite{borkar08stochastic-approximation-book}) based update rules for $\lambda(t)$. But the above remark tells us that the optimal solution of the constrained problem~\eqref{eqn:centralized-constrained-problem} requires us to randomize between two values of $\lambda$ in case the optimal $\lambda^*$ as in 
Theorem~\ref{theorem:relation-between-constrained-and-unconstrained-problems} belongs to the set of such discontinuities. However, this randomization will require us to update a randomization probability at another timescale; having stochastic approximations running in multiple timescales leads to  slow convergence. Hence, instead of using a varying $\beta(t)$, we use a fixed, but large $\beta$ and update $\lambda(t)$ in an iterative fashion using stochastic approximation. Our proposed  Algorithm~\ref{algorithm:gibbs-learning-algorithm-for-constrained-problem} updates $\lambda(t)$ iteratively in order to  solve \eqref{eqn:centralized-constrained-problem}.

\noindent\fbox{
    \parbox{0.46 \textwidth}{
\begin{algorithm}
 Choose any initial $B(0) \in \{0,1\}^N$ and $\lambda(0) \geq 0$. 
At each discrete time instant $t=0,1,2,\cdots$, pick a random sensor $j_t \in \mathcal{N}$ independently and uniformly. For sensor $j_t$, choose $B_{j_t}(t)=1$ with probability 
$p:=\frac{ e^{-\beta h_{\lambda(t)}(B_{-j_t}(t-1),1)}}{e^{-\beta h_{\lambda(t)}(B_{-j_t}(t-1),1)}+e^{-\beta h_{\lambda(t)}(B_{-j_t}(t-1),0)}}$ and choose $B_{j_t}(t)=0$ with probability 
$(1-p)$. For $k  \neq j_t$, we choose $B_k(t)=B_k(t-1)$. 

After this operation, before the $(t+1)$ decision instant, update  $\lambda(t)$ at each node as follows. 

$$\lambda(t+1)=[\lambda(t)+a(t) (||B(t-1)||_1-\bar{N})]_b^c$$
The stepsize $\{a(t)\}_{t \geq 1}$ constitutes a positive sequence such that $\sum_{t=1}^{\infty}a(t)=\infty$  and $\sum_{t=1}^{\infty}a^2(t)=\infty$. The nonnegative projection boundaries $b$ and $c$ for the $\lambda(t)$ iterates are such that  $\lambda^* \in (b,c)$ where $\lambda^*$ is defined in Assumption~\ref{assumption:existence-of-optimal-lambda}.
\label{algorithm:gibbs-learning-algorithm-for-constrained-problem}
\vspace{2mm}
 \end{algorithm}
 }}  

The update of $\lambda(t)$ in Algorithm~\ref{algorithm:gibbs-learning-algorithm-for-constrained-problem} is inspired by the following result which is crucial in the convergence proof.
\noindent\fbox{
    \parbox{0.46 \textwidth}{
\begin{lemma}\label{lemma:active-sensors-decreasing-in-lambda-under-basic-gibbs-sampling}
Under Algorithm~\ref{algorithm:basic-gibbs-for-known-distribution-iid-data}, 
$\mathbb{E}_{\mu_2} ||B(t)||_1$ is a Lipschitz continuous and decreasing function of $\lambda$.
\end{lemma}
}}
\begin{proof}
See Appendix~\ref{appendix:proof-of-active-sensors-decreasing-in-lambda-under-basic-gibbs-sampling}.
\end{proof}

{\em Discussion of Algorithm~\ref{algorithm:gibbs-learning-algorithm-for-constrained-problem}:}
\begin{itemize}
\item If $||B(t-1)||_1$ is more than $\bar{N}$, then $\lambda(t)$ is increased with the hope that this will reduce the number of active sensors in subsequent iterations, as suggested by Lemma~\ref{lemma:active-sensors-decreasing-in-lambda-under-basic-gibbs-sampling}.
\item The $B(t)$ and $\lambda(t)$ processes run on two different timescales; $B(t)$ runs in the faster timescale whereas $\lambda(t)$ runs in a slower timescale. This can be understood from the fact that the stepsize in the $\lambda(t)$ update process decreases with time $t$. Here the faster timescale iterate will view the slower timescale iterate as quasi-static, while the slower timescale iterate will view the faster timescale as almost equilibriated. This is reminiscent of  two-timescale stochastic approximation (see \cite[Chapter~$6$]{borkar08stochastic-approximation-book}).
\end{itemize}

We make the following feasibility assumption for \eqref{eqn:centralized-constrained-problem}, under the chosen $\beta>0$. 
\begin{assumption}\label{assumption:existence-of-optimal-lambda}
There exists $\lambda^* \geq 0$ such that the constraint in \eqref{eqn:centralized-constrained-problem} under 
$\lambda^*$ and Algorithm~\ref{algorithm:basic-gibbs-for-known-distribution-iid-data} is met with equality.
\end{assumption}
\begin{remark}
By Lemma~\ref{lemma:active-sensors-decreasing-in-lambda-under-basic-gibbs-sampling}, $\mathbb{E}||B||_1$ continuously decreases in $\lambda$. Hence, if $\bar{N}$ is feasible, then such a $\lambda^*$ must exist by the {\em intermediate value theorem}.
\end{remark}
Let us define:
$h_{\lambda(t)}(B):=f(B)+\lambda(t) ||B||_1$. Let $\pi_{\beta| \lambda^*}(\cdot)$ denote  $\pi_{\beta}(\cdot)$ under $\lambda=\lambda^*$.

\noindent\fbox{
    \parbox{0.46 \textwidth}{
\begin{theorem}\label{theorem:optimality-of-the-learning-algorithm-for-constrained-problem}
Under Algorithm~\ref{algorithm:gibbs-learning-algorithm-for-constrained-problem} and Assumption~\ref{assumption:existence-of-optimal-lambda}, we have $\lambda(t) \rightarrow \lambda^*$ almost surely, and the limiting distribution of $\{B(t)\}_{t \geq 0}$ is $\pi_{\beta| \lambda^*}(\cdot)$.
\end{theorem}
}}
\begin{proof}
See Appendix~\ref{appendix:proof-of-optimality-of-the-learning-algorithm-for-constrained-problem}.
\end{proof}
This theorem says that Algorithm~\ref{algorithm:gibbs-learning-algorithm-for-constrained-problem}   produces a configuration from the distribution $\pi_{\beta| \lambda^*}(\cdot)$ under steady state.

\subsection{A hard constraint on the number of activated sensors}\label{subsection:hard-constraint}
Let us consider the following modified constrained problem:
\begin{equation}\label{eqn:constrained-optimization-problem-static-data-parametric-distribution}
\min_{B \in \mathcal{B}} f(B) \textbf{ s.t. } ||B||_1 \leq \bar{N} 
\tag{MCP}
\end{equation}
It is easy to see that \eqref{eqn:constrained-optimization-problem-static-data-parametric-distribution} can be easily solved using similar Gibbs sampling algorithms as in Section~\ref{section:gibbs-sampling-unconstrained-problem}, where the Gibbs sampling algorithm runs only on the set of configurations which activate $\bar{N}$ number of sensors. 
Thus, as a by-product, we have also proposed a methodology for the problem in \cite{wang-etal16efficient-observation-selection}, though our framework is more general than   \cite{wang-etal16efficient-observation-selection}. 

\begin{remark}
The constraint in \eqref{eqn:centralized-constrained-problem} is weaker than  \eqref{eqn:constrained-optimization-problem-static-data-parametric-distribution}. 
\end{remark}
\begin{remark}
If we choose $\beta$ very large, then the number of sensors activated by GIBBSLEARNING will have very small variance. This allows us to solve \eqref{eqn:constrained-optimization-problem-static-data-parametric-distribution} with high probability.
\end{remark}

\section{Centralized Tracking of  IID   process: Unknown $\theta_0$}
\label{section:iid-data}
In Section~\ref{section:gibbs-sampling-unconstrained-problem}, we described algorithms for centralized tracking of an i.i.d. process where $p_{\theta_0}(\cdot)$ is known. In this section, we will deal with the centralized tracking of an i.i.d. process $\{X(t)\}_{t \geq 0}$ where $X(t) \sim p_{\theta_0}(\cdot)$ with $\theta_0$ unknown; in this case, $\theta_0$ has to be learnt over time through observations, which creates  many nontrivial issues that need to be addressed before using Gibbs sampling for sensor subset selection.
\subsection{The proposed algorithm for unknown $\theta_0$}\label{subsubsection:GIBBSLEARNINGEM-algorithm}
Since $\theta_0$ is unknown,   its estimate $\theta(t)$ has to be updated over time using the sensor observations. On the other hand, to solve the constrained problem \eqref{eqn:centralized-constrained-problem}, we need to update $\lambda(t)$ over time so as to attain the optimal $\lambda^*$ of Theorem~\ref{theorem:relation-between-constrained-and-unconstrained-problems} iteratively. Further, $f(B)$ (MSE under configuration $B$) in \eqref{eqn:centralized-unconstrained-problem-with-f-and-h} is unknown since $\theta_0$ is unknown, and $f^{(t)}(B)$ has to be learnt over time using the sensor observations.  Hence, we combine the Gibbs sampling algorithm with update schemes for $f^{(t)}(B)$, $\lambda(t)$ and $\theta(t)$  using stochastic approximation (see \cite{borkar08stochastic-approximation-book}).

 The algorithm also requires a sufficiently large positive number $A_0$ and a large integer $T$ as input. 

Let $\mathcal{J}(t)$ denote the indicator that time $t$ is an integer multiple of $T$.  Define $\nu(t):=\sum_{\tau=0}^t \mathcal{J}(\tau)$. 

We first describe some key features and steps of the algorithm and then provide a brief summary of the algorithm.
\subsubsection{Step size}\label{subsubsection:step-size-iid-centralized}
 For the stochastic approximation updates, the algorithm uses step sizes which are nonnegative sequences $\{a(t)\}_{t \geq 0}$, $\{b(t)\}_{t \geq 0}$, $\{c(t)\}_{t \geq 0}$, $\{d(t)\}_{t \geq 0}$ such that:

\footnotesize
(i) $\sum_{t=0}^{\infty}a(t)=\sum_{t=0}^{\infty}b(t)=\sum_{t=0}^{\infty}c(t)=\infty ,$
(ii) $\sum_{t=0}^{\infty}a^2(t)<\infty, \sum_{t=0}^{\infty}b^2(t)<\infty, \sum_{t=0}^{\infty}c^2(t)<\infty ,$
(iii) $\lim_{t \rightarrow \infty} d(t)=0$, 
(iv) $\sum_{t=0}^{\infty}\frac{c^2(t)}{d^2(t)}<\infty$,
(v) $\lim_{t \rightarrow \infty}\frac{b(t)}{a(t)}=\lim_{t \rightarrow \infty}\frac{c(\lfloor \frac{t}{T} \rfloor )}{b(t)}=0$.\\
\normalsize
\subsubsection{Gibbs sampling step}\label{subsubsection:gibbs-sampling-step-iid-centralized}
The algorithm also maintains a running estimate $h^{(t)}(B)$ of $h(B)$ for all $B \in \mathcal{B}$. 
At time~$t$, it selects a random sensor $j_t \in \mathcal{N}$ with uniformly and independently, and sets  $B_{j_t}(t)=1$ with probability 
$p(t):=\frac{ e^{-\beta h^{(t)}(B_{-j_t}(t-1),1)}}{e^{-\beta h^{(t)}(B_{-j_t}(t-1),1)}+e^{-\beta h^{(t)}(B_{-j_t}(t-1),0)}}$ and  $B_{j_t}(t)=0$ with probability 
$(1-p(t))$ (similar to Algorithm~\ref{algorithm:basic-gibbs-for-known-distribution-iid-data}). For $k  \neq j_t$, it sets $B_k(t)=B_k(t-1)$. This operation can even be repeated multiple times. The sensors are activated according to   $B(t)$, and the observations  $Z_{B(t)}(t):=\{z_k(t):B_k(t)=1\}$ are collected. Then the algorithm declares $\hat{X}(t)=\mu_2(\mathcal{H}_c(t))$. 

\subsubsection{Occasional reading of all the sensors}
If $\mathcal{J}(t)=1$, the fusion center reads all sensors and obtains $Z(t)$. This is required primarily because we seek to update $\theta(t)$ iteratively and reach a local maximum of the function  $g(\theta)= \mathbb{E}_{X(t) \sim p_{\theta_0}(\cdot), B(t)=[1,1,\cdots,1]}\log p (Z(t) | \theta)$.

\subsubsection{$\theta(t)$ update when $\mathcal{J}(t)=1$}
Since we seek to reach a local maximum of $g(\theta)= \mathbb{E}_{X(t) \sim p_{\theta_0}(\cdot), B(t)=[1,1,\cdots,1]}\log p (Z(t) | \theta)$, a gradient ascent scheme needs to be used. The gradient of $g(\theta)$ along any coordinate can be computed by perturbing $\theta$ in two opposite directions along that coordinate and evaluating the difference of  $g(\cdot)$ at those two perturbed values.  However, if $\theta_0$ is high-dimensional, then estimating this gradient along each coordinate is computationally intensive. Moreover, evaluating $g(\theta)$ for any $\theta$ requires us to compute an expectation, which might also be expensive. Hence, we perform a noisy gradient estimation for   $g(\theta)$ by simultaneous perturbation stochastic approximation (SPSA) as in \cite{spall92original-SPSA}. Our algorithm generates 
$\Delta(t) \in \{1,-1\}^d$ uniformly over all sequences, and perturbs  the current estimate $\theta(t)$ by a random vector $d(\nu(t)) \Delta(t)$  in two opposite directions, and estimates each component of the gradient from the difference $\log p(Z(t)|  \theta(t)+d(\nu(t)) \Delta(t)  ) -\log p(Z(t)|  \theta(t)-d(\nu(t)) \Delta(t)  ) $;   this estimate is noisy because (i) $Z(t)$ and $\Delta(t)$ are random, and (ii) $d(\nu(t))>0$.   

The $k$-th component of $\theta(t)$ is updated as follows:

\footnotesize
\begin{eqnarray}\label{eqn:theta-update-iid-data}
\theta_k(t+1) 
&=& \bigg[ \theta_k(t)+c(\nu(t)) \mathcal{J}(t)  \bigg( \frac{  \log p(Z(t)|  \theta(t)+d(\nu(t)) \Delta(t)  )   }{2 d(\nu(t)) \Delta_k(t)} \nonumber\\
&& - \frac{  \log p(Z(t)|   \theta(t)-d(\nu(t)) \Delta(t)  )   }{2 d(\nu(t)) \Delta_k(t)} \bigg)  \bigg]_{\Theta}
\end{eqnarray}
\normalsize
The iterates are projected onto the compact set $\Theta$ to ensure boundedness. The diminishing sequence $\{d(t)\}_{t \geq 0}$ ensures that the gradient estimate becomes more accurate with time.

\subsubsection{$\lambda(t)$ update}
$\lambda(t)$ is updated as follows:
\begin{equation}\label{eqn:lambda-update-iid-data}
\lambda(t+1)=[\lambda(t)+b(t) (||B(t)||_1-\bar{N})]_0^{A_0}.
\end{equation}
The intuition here is that, if $||B(t)||_1>\bar{N}$, the sensor activation cost $\lambda(t)$ needs to be increased to prohibit activating large number of sensors in future; this is motivated by Lemma~\ref{lemma:active-sensors-decreasing-in-lambda-under-basic-gibbs-sampling}. The goal is to converge to $\lambda^*$ as defined in Theorem~\ref{theorem:relation-between-constrained-and-unconstrained-problems}.

\subsubsection{$f^{(t)}(B)$ update}
Since $p_{\theta_0}(X)$ is not known initially, the true value of $f(B)$   is not known;  hence, the algorithm updates an estimate $f^{(t)}(B)$ using the sensor observations. 
If $\mathcal{J}(t)=1$, the fusion center obtains $Z(t)$ by reading all sensors. The goal is to obtain an estimate $Y_B(t)$ of the MSE under a configuration $B$, by using these observations, and update $f^{(t)}(B)$ using $Y_B(t)$. 
However, since $\theta_0$ is unknown and only $\theta(t)$ is available, as an alternative to the MSE under configuration $B$, the fusion center uses the trace of the conditional covariance matrix of $X(t)$ given $Z_B(t)$,  assuming that $X(t) \sim p_{\theta(t)}(\cdot)$. Hence, we define a random variable  $Y_B(t):= \mathbb{E}_{X(t) \sim p(\cdot|\theta(t),Z_B(t))}(||X(t)-\hat{X}_B(t)||^2|Z_B(t), \theta(t)) $ for all $B \in \mathcal{B}$, where    $\hat{X}_B(t)$ is the MMSE estimate declared by $\mu_2$ under configuration $B$, and $Z_B(t)$ is the observation made by active sensors determined by $B$. Clearly, $Y_B(t)$ is a random variable with the randomness coming from two sources: (i) randomness of $\theta(t)$, and (ii) randomness of $Z_B(t)$ which has a distribution $p(Z_B(t)|\theta_0)$ since the original $X(t)$ process that yields $Z_B(t)$ has a distribution $p_{\theta_0}(\cdot)$. Computation of $Y_B(t)$ is simple for Gaussian $X(t)$ and the MMSE estimator, since closed form expressions are available to compute $Y_B(t)$. 

Using $Y_B(t)$, the following   update is made for all $B \in \mathcal{B}$:

\footnotesize
\begin{eqnarray} \label{eqn:fB-update-iid-data}
f^{(t+1)}(B)= [f^{(t)}(B)+\mathcal{J}(t)  a(\nu(t))   (Y_B(t) - f^{(t)}(B)) ]_0^{A_0}
\end{eqnarray}
\normalsize
The iterates are projected onto   $[0,A_0]$ to ensure boundedness. The goal here is that, if  $\theta(t) \rightarrow \theta^*$, then $f^{(t)}(B)$ will converge to $\mathbb{E}_{Z_B(t) \sim p(\cdot| \theta_0)} \mathbb{E}_{X(t) \sim p(\cdot|\theta^*,Z_B(t))}(||X(t)-\hat{X}_B(t)||^2|Z_B(t), \theta^*) $, which is equal to  $f(B)$ under $\theta^*=\theta_0$.

{\em We will later argue that this occasional $O(2^N)$ computation for all $B \in \mathcal{B}$ can be avoided, but   convergence   will be slow.}

\subsubsection{The algorithm} A summary of all the steps of our scheme is provided in Algorithm~\ref{algorithm:gibbs-learning-algorithm-for-constrained-problem-EM}.   {\em We will show in Theorem~\ref{theorem:convergence-of-GIBBSLEARNINGEM} that this algorithm almost surely converges to the set of locally optimum solutions for \eqref{eqn:centralized-constrained-problem}.}

\noindent\fbox{
    \parbox{0.46 \textwidth}{
\begin{algorithm}
 Initialize all iterates arbitrarily.\\
{\bf For any  time  $t=0,1,2,\cdots$:}\\
1. Perform the Gibbs sampling step as in Section~\ref{subsubsection:gibbs-sampling-step-iid-centralized}, obtain the observations $Z_{B(t)}(t)$, and estimate $\hat{X}(t)$. Update $\lambda(t)$ according to \eqref{eqn:lambda-update-iid-data}.

2. If $\mathcal{J}(t)=1$, compute $h^{(t)}(B)=f^{(t)}(B)+\lambda(t) ||B||_1$ for all $B \in \mathcal{B}$. Read all sensors and obtain $Z(t)$. Update $f^{(t)}(B)$ for all $B \in \mathcal{B}$ using \eqref{eqn:fB-update-iid-data} and $\theta(t)$ using \eqref{eqn:theta-update-iid-data}.

\label{algorithm:gibbs-learning-algorithm-for-constrained-problem-EM}
 \end{algorithm}  
}}

\subsubsection{Multiple timescales in Algorithm~\ref{algorithm:gibbs-learning-algorithm-for-constrained-problem-EM}} Algorithm~\ref{algorithm:gibbs-learning-algorithm-for-constrained-problem-EM} has multiple iterations running in multiple timescales (see \cite[Chapter~$6$]{borkar08stochastic-approximation-book}). The $\{B(t)\}_{t \geq 0}$ process runs ar the fastest timescale, whereas the $\{ \theta(t)\}_{t \geq 0}$ update scheme runs at the slowest timescale. The basic idea is that a faster timescale iterate views a slower timescale iterate as quasi-static, whereas a slower timescale iterate views a faster timescale iterate as almost equilibriated. For example, since $\lim_{t \rightarrow \infty}\frac{c(t)}{a(t)}=0$, the $\theta(t)$ iterates will vary very slowly compared to $f^{(t)}(B)$ iterates; as a result, $f^{(t)}(B)$ iterates will view quasi-static $\theta(t)$.

\subsection{Complexity of Algorithm~\ref{algorithm:gibbs-learning-algorithm-for-constrained-problem-EM}}
\subsubsection{Sampling and communication complexity}
Since all   sensors are activated when $\mathcal{J}(t)=1$, the mean number of additional active sensors   per unit time is $O(\frac{N}{T})$; these observations  need to be communicated to the fusion center.  {\bf $O(\frac{N}{T})$ can be made arbitrarily small by choosing $T$ large enough. }

\subsubsection{Computational complexity}\label{subsubsection:computational-complexity-of-centralized-iid-tracking}
{\em The computation of $Y_B(t)$ in \eqref{eqn:fB-update-iid-data} for all $B \in \mathcal{B}$ requires $O(2^N)$ computations whenever $\mathcal{J}(t)=1$. However, if one chooses large  $T$ (e.g., $O(4^N)$), then this additional  computation per unit time will be small. However, if one wants to avoid that computation also, then one can simply compute $Y_{B(t)}(t)$ and update $f^{(t)}(B(t))$ instead of doing it for all configurations $B \in \mathcal{B}$. However,  the stepsize sequence $a(\nu(t))$ cannot be used; instead, a stepsize $a(\nu_B(t))$ has to be used when $B(t)=B$ and $f^{(t)}(B)$ is updated using \eqref{eqn:fB-update-iid-data}, where $\nu_B(t):=\sum_{\tau=0}^t \mathcal{J}(\tau) \mathbb{I}(B(\tau)=B)$ is the number of times configuration $B$ was chosen till time $t$.  In this case, the convergence result (Theorem~\ref{theorem:convergence-of-GIBBSLEARNINGEM}) on Algorithm~\ref{algorithm:gibbs-learning-algorithm-for-constrained-problem-EM} will still hold; however, the proof will require a technical condition $\lim \inf_{t \rightarrow \infty} \frac{\nu_B(t)}{t}>0$ almost surely for all $B \in \mathcal{B}$, which will be satisfied by the Gibbs sampler using finite $\beta$ and bounded $h^{(t)}(B)$. However, we discuss only \eqref{eqn:fB-update-iid-data} update in this paper for the sake of simplicity in the convergence proof, since technical details of asynchrounous stochastic approximation required in the variant is not the main theme of this paper.}

{\em 
When $\mathcal{J}(t)=1$, one can avoid computation of $h^{(t+1)}(B)$ for all $B \in \mathcal{B}$  in Step~$2$ of Algorithm~\ref{algorithm:gibbs-learning-algorithm-for-constrained-problem-EM}. Instead, the fusion center can update only  $h^{(t)}(B(t))$, $h^{(t)}(B_{-j_t}(t-1),1)$ and $h^{(t)}(B_{-j_t}(t-1),0)$ at time $t$, since only these iterates are  required in the Gibbs sampling.}

\subsection{Convergence of Algorithm~\ref{algorithm:gibbs-learning-algorithm-for-constrained-problem-EM}}

\subsubsection{List of assumptions}\label{subsubsection:list-of-assumptions}
\begin{assumption}\label{assumption:Lipschitz-continuity-wrt-theta}
The distribution $p_{\theta}(\cdot)$ and the mapping $\mu_2(\cdot; \cdot; \theta)$ (or $\mu_2^{(k)}$ for distributed case) as defined before are Lipschitz continuous in $\theta \in \Theta$.
\end{assumption}

\begin{assumption}\label{assumption:fixed-decoding-strategy-for-iid-data}
$\mu_2$ is known to the fusion center (centralized case), and $\{\mu_2^{(k)}\}_{k \geq 1}$ are known  to all sensors (distributed). 
\end{assumption}

\begin{remark}
Assumption~\ref{assumption:fixed-decoding-strategy-for-iid-data} allows us to focus only on the sensor subset selection problem rather than the problem of estimating the process given the sensor observations. For optimal MMSE estimators, $\mu_2(\mathcal{H}_c(t))=\mathbb{E}(X(t)|\mathcal{H}_c(t))$. Computation of $\mu_2(\cdot)$ will depend on the exact functional form of $p_{\theta}(X)$, and it can be done by using Bayes' theorem.
\end{remark}

\begin{assumption}\label{assumption:existence-of-lambda*}
Let us consider $Y_B(t)$ with $\theta(t)=\theta$ fixed in Algorithm~\ref{algorithm:gibbs-learning-algorithm-for-constrained-problem-EM}. Suppose that, one uses Algorithm~\ref{algorithm:basic-gibbs-for-known-distribution-iid-data} to solve \eqref{eqn:centralized-unconstrained-problem} for a given $\lambda$ but with the MSE $||X(t)-\hat{X}(t)||^2$ replaced by $Y_{B(t)}(t)$ in the objective function of \eqref{eqn:centralized-unconstrained-problem}, and then finds the $\lambda^*(\theta)$ as in Theorem~\ref{theorem:relation-between-constrained-and-unconstrained-problems} to meet the constraint $\bar{N}$. We assume that, 
for the given $\beta$ and $\bar{N}$, and for each $\theta \in \Theta$, there exists $\lambda^*(\theta) \in [0,A_0)$ such that,  the optimal Lagrange multiplier to relax this new unconstrained problem is $\lambda^*(\theta)$ (Theorem~\ref{theorem:relation-between-constrained-and-unconstrained-problems}). Also, $\lambda^*(\theta)$ is Lipschitz continuous in $\theta \in \Theta$.
\end{assumption}
\begin{remark}
Assumption~\ref{assumption:existence-of-lambda*} makes sure that the $\lambda(t)$ iteration \eqref{eqn:lambda-update-iid-data} converges, and the   constraint is met with equality.
\end{remark}
Let us define the function $\bar{\Gamma}_{\theta}(\phi):=\lim_{\delta \downarrow 0} \frac{[\theta+\delta \phi]_{\Theta}-\theta}{\delta}$.
\begin{assumption}\label{assumption:finite-number-of-local-maximum}
Consider the  function $g(\theta)= \mathbb{E}_{X(t) \sim p_{\theta_0}(\cdot), B(t)=[1,1,\cdots,1]}\log p (Z(t) | \theta)$; this is the expected conditional log-likelihood function of $Z(t)$ conditioned on $\theta$, given that $X(t) \sim p_{\theta_0}(\cdot)$ and $B(t)=[1,1,\cdots,1]$. We assume    that the ordinary differential equation $\dot{\theta}(t)=\bar
{\Gamma}_{\theta(t)}(\nabla g(\theta(t)))$ has a globally  asymptotically stable solution $\theta^*$ in the interior of $\Theta$. Also,  $\nabla g(\theta)$ is Lipschitz continuous in $\theta$.
\end{assumption}
\begin{remark}
One can show that the $\theta(t)$ iteration~\eqref{eqn:theta-update-iid-data} asymptotically tracks the ordinary differential equation $\dot{\theta}(\tau)=\nabla g(\theta(\tau))$ inside the interior of $\Theta$. In fact, $\bar{\Gamma}_{\theta(\tau)}(\nabla g(\theta(\tau))=\nabla g(\theta(\tau))$ when $\theta(\tau)$ lies inside the interior of $\Theta$. The condition on $\dot{\theta}(\tau)=\bar
{\Gamma}_{\theta(\tau)}( \nabla g(\theta(\tau)))$ is required to make sure that the iteration does not converge to some unwanted point on the boundary of $\Theta$ due to the forced projection. The assumption on $\theta^*$ makes sure that the $\theta(t)$ iteration converges to $\theta^*$. 
\end{remark}

\subsubsection{The main result}
Now we present    convergence result for Algorithm~\ref{algorithm:gibbs-learning-algorithm-for-constrained-problem-EM}. This result tells us that {\em the iterates of Algorithm~\ref{algorithm:gibbs-learning-algorithm-for-constrained-problem-EM} almost surely converge to the set of local optima for   \eqref{eqn:centralized-constrained-problem}.}
\noindent\fbox{
    \parbox{0.46 \textwidth}{
\begin{theorem}\label{theorem:convergence-of-GIBBSLEARNINGEM}
Under Assumptions~\ref{assumption:Lipschitz-continuity-wrt-theta},  \ref{assumption:fixed-decoding-strategy-for-iid-data},  \ref{assumption:existence-of-lambda*},  \ref{assumption:finite-number-of-local-maximum} and Algorithm~\ref{algorithm:gibbs-learning-algorithm-for-constrained-problem-EM}, we have  $\lim_{t \rightarrow \infty}\theta(t)=\theta^*$ almost surely. Correspondingly, $\lambda(t) \rightarrow \lambda^*(\theta^*)$ almost surely. 
Also, $f^{(t)}(B) \rightarrow\mathbb{E}_{Z_B(t) \sim p(\cdot| \theta_0)} \mathbb{E}_{X(t) \sim p(\cdot|\theta^*,Z_B(t))}(||X(t)-\hat{X}_B(t)||^2|Z_B(t), \theta^*) =:f_{\theta^*}(B)$ almost surely for all $B \in \mathcal{B}$.  The $B(t)$ process reaches the steady-state distribution  $\pi_{\beta,f_{\theta^*},\lambda^*(\theta^*), \theta^*}(\cdot)$ which can be obtained by replacing $h(B)$ in \eqref{eqn:definition-of-Gibbs-distribution} by $f_{\theta^*}(B)+\lambda^*||B||_1$.
\end{theorem}
}}
\begin{proof}
See Appendix~\ref{appendix:proof-of-GIBBSLEARNINGEM}.
\end{proof}

\begin{remark} 
If $\theta(t) \rightarrow \theta^*$, but the constraint in \eqref{eqn:centralized-constrained-problem} is satisfied  with $\lambda=0$ and  policy $\mu_2(\cdot; \cdot; \theta^*)$, then  $\lambda(t) \rightarrow 0$, i.e., $\lambda^*(\theta^*)=0$, and the constraint becomes redundant. {\em If $\theta^*=\theta_0$, then the algorithm reaches the global optimum.}
\end{remark}
\begin{remark}\label{remark:what-if-all-sensors-are-not-read}
If all sensors are not read when $\mathcal{J}(t)=1$, then one has to update $\theta(t)$ based on the observations   $Z_{B(t)}(t)$ collected from the sensors determined by $B(t)$. In that case, $\theta(t)$ will converge to a local maximum $\theta_1$ of $\lim_{t \rightarrow \infty}\mathbb{E}_{\pi_{\beta, f_{\theta}, \lambda^*(\theta), \theta}}  \log p(Z_{B(t)}(t)|\theta)$, which will be different from $\theta^*$ of Theorem~\ref{theorem:convergence-of-GIBBSLEARNINGEM} in general. However, in the numerical example   in Section~\ref{section:numerical-results}, we observe that $\theta_1=\theta^*$.
\end{remark}

\section{Distributed tracking of the i.i.d. process}
\label{section:iid-data-distributed-estimation}
We next seek to solve the constrained problem~\eqref{eqn:distributed-constrained-problem}. This problem brings in additional challenges compared to \eqref{eqn:centralized-constrained-problem}: (i) each sensor~$k$ has access only  to its local measurement, i.e., $z_k(t)$,  if $B_k(t)=1$, (ii) sharing measurements   across the network will consume a large amount of energy and bandwidth, and (iii) ideally, the iterates $\lambda(t)$ and $\theta(t)$ should be known at all sensors. To resolve these issues, we propose an algorithm  that  combines Algorithm~\ref{algorithm:gibbs-learning-algorithm-for-constrained-problem-EM} with consensus among sensor nodes (see \cite{xiao-etal05scheme-sensor-fusion-average-consensus}). However, our approach is different from traditional consensus schemes in the following aspects: (i) traditional consensus schemes run many steps of a consensus iteration, thus requiring many rounds of message exchange among neighbouring nodes, and (ii) traditional consensus schemes do not care about the correctness of the data at any particular sensor node. In contrast,  our proposed algorithm  allows each sensor to broadcast its local information only once to its neighbours in a time slot. Also, since many of the sensors may use outdated estimates, we propose an on-line learning scheme based on stochastic approximation in order to optimize the coefficients of the linear combination used in consensus.

\subsection{The proposed algorithm}
\label{subsection:algorithm-for-distributed-estimation-iid-data}

Note that,  {\em if   the Gibbs sampling step of Algorithm~\ref{algorithm:gibbs-learning-algorithm-for-constrained-problem-EM} is run at all sensors to make their individual activation decisions, but all sensors are supplied with the same initial seed for randomization, then all sensors will sample the same $B(t)$ at each time $t$. }  We will exploit this fact in the next algorithm. However, depending on the current configuration $B(t)=B$, each node uses a linear combination of its own estimate and the estimates made by its neighbours. Let us denote the {\em initial} estimate made by node~$k$ at time~$t$ by $\bar{X}^{(k)}(t)$; this estimation is done at node~$k$ based on $\mathcal{H}_c(t)$. The actual estimate $\hat{\mathbf{X}}(t):=[\hat{X}^{(1)}(t), \hat{X}^{(2)}(t), \cdots, \hat{X}^{(N)}(t)]'$ is obtained from  $\bar{\mathbf{X}}(t):=[\bar{X}^{(1)}(t), \bar{X}^{(2)}(t), \cdots, \bar{X}^{(N)}(t)]'$
 by $\hat{\mathbf{X}}(t)= K_{B(t)}^{(t)} \bar{\mathbf{X}}(t) $; this method is motivated by the Kalman consensus filter proposed in  \cite{saber09kalman-consensus-optimality-stability}. Here $K_B^{(t)} \in \mathbb{R}^{N \times N}$ is the weight matrix to be used at time $t$ under configuration $B$; this matrix has $(i,j)$-th entry zero if nodes $i$ and $j$ are not connected in the wireless network, it can induce a consensus such as in \cite{xiao-etal05scheme-sensor-fusion-average-consensus}. The matrices $K_B^{(t)}$ for all $B \in \mathcal{B}$ are updated for all $t$ when  $\mathcal{J}(t)=1$, and  are broadcast to the sensors. As with the $\theta(t)$ update  in \eqref{eqn:theta-update-iid-data},  we use SPSA  to find optimal $K_B$  in order to minimize the  error.
 
 First we describe some special steps of the algorithm. Define $\nu(t)=\sum_{\tau=1}^t \mathcal{J}(\tau)$, and $\nu_B(t)=\sum_{\tau=1}^t \mathcal{J}(\tau) \mathbb{I}(B(\tau)=B)$ (the number of times configuration $B$ is sampled till time $t$).
 
 \subsubsection{$f^{(t)}(B)$ update when $\mathcal{J}(t)=1$}\label{subsubsection:Y_K_B-and-f-t-B-update}
If $\mathcal{J}(t)=1$, all sensors are read to obtain $Z(t)$; this $Z(t)$ is either supplied to all sensors, or sent to some specific node which does centralized computation (this is done only when $\mathcal{J}(t)=1$) and broadcasts the results to all sensors. For each $B \in \mathcal{B}$,  all estimates $\{\hat{X}_B^{(k)}(t)\}_{1 \leq k \leq N}$ are computed using  $Z_B(t)$, where $\hat{X}_B^{(k)}(t)$ denotes the estimate at node~$k$ at time $t$ if $B(t)=B$. Then, the quantity   
$ Y_{K_B^{(t)}}:=\mathbb{E}_{X(t) \sim p(\cdot|\theta(t),Z_B(t), K_B^{(t)})} (\frac{1}{N}\sum_{k=1}^N  ||X(t)-\hat{X}_B^{(k)}(t)||^2| \theta(t),Z_B(t), K_B^{(t)})$ is computed, and  the following update is done for all $B \in \mathcal{B}$:

\footnotesize
\begin{eqnarray} \label{eqn:fB-update-iid-data-distributed}
f^{(t+1)}(B)= f^{(t)}(B)+\mathcal{J}(t)  a(\nu(t))     \bigg(Y_{K_B^{(t)}} - f^{(t)}(B) \bigg) 
\end{eqnarray}
\normalsize
We will see in Section~\ref{subsubsection:complexity-distributed-iid} that this $O(2^N)$ computation can be avoided without sacrificing convergence.
\subsubsection{$K_B^{(t)}$ update when $\mathcal{J}(t)=1$} The algorithm requires another condition $\sum_{t=0}^{\infty} \frac{b^2(t)}{d^2(t)} < \infty$ (apart from those in Section~\ref{subsubsection:step-size-iid-centralized}) to ensure convergence of the $K_B^{(t)}$ iterates which are updated via the following SPSA algorithm whenever $\mathcal{J}(t)=1$ (i.e., $t$ is an integer multiple of $T$). 

A random matrix  $\Gamma_t \in \mathbb{R}^{ N \times N}$ is generated such that $\Gamma_t(i,j)=0$ if  $i$ and $j$ are not neighbours, otherwise $\Gamma_t (i,j) \in \{-1,1\}$  independently with equal probability. Now,  the following updates are done for all links $(i,j)$:
 \footnotesize
\begin{eqnarray}\label{eqn:coefficients-update-iid-data-distributed}
&& K_B^{(t+1)}(i,j) =
 \bigg[ K_B^{(t)}(i,j)-b(\nu_B(t)) \mathcal{J}(t)  \nonumber \\
 && \mathbb{I}(B(t)=B)  \bigg( \frac{  Y_{K_{B}^{(t)}+d(\nu_B(t)) \Gamma_t}   }{2 d(\nu_B(t)) \Gamma_t(i,j)}  - \frac{  Y_{K_{B}^{(t)}-d(\nu_B(t)) \Gamma_t}  }{  2 d(\nu_B(t)) \Gamma_t(i,j)  } \bigg)  \bigg]_{-A_0}^{A_0}
\end{eqnarray}
\normalsize
\eqref{eqn:coefficients-update-iid-data-distributed} is a gradient descent scheme, with the goal to converge to some $K_B^*$  so that $\lim_{t \rightarrow \infty}\mathbb{E}(Y_{K_B^{(t)}}(t))$ (if it exists) is minimized.   $\nabla \mathbb{E}(Y_{K_B^{(t)}}(t))$ is estimated by SPSA.

\subsubsection{Outline of the proposed algorithm}
The entire scheme   is described in Algorithm~\ref{algorithm:distributed-gibbs-learning-algorithm-for-constrained-problem-EM}.  

 \noindent\fbox{
    \parbox{0.46 \textwidth}{
 \begin{algorithm}
The same seed is supplied to all sensors for Gibbs sampling. All iterates are initialized arbitrarily.

{\bf For any time $t=0,1,2,\cdots$, do at each sensor~$k$:}

1. Run   Gibbs sampling   as in Section~\ref{subsubsection:gibbs-sampling-step-iid-centralized}, once or multiple times. All sensors will have the same $B(t)$; they will be activated according to $B(t)$ and make observations.  Compute $\bar{X}^{(k)}(t)=\mu_2^{(k)}(\mathcal{H}_c(t))$ and $\hat{\mathbf{X}}(t)=K_{B(t)}^{(t)} \bar{\mathbf{X}}(t)$ locally. Update  $\lambda(t)$ using \eqref{eqn:lambda-update-iid-data}.  

2.   If $\mathcal{J}(t)=1$, read all sensors and obtain $Z(t)$. For each $B \in \mathcal{B}$, compute all estimates $\{\hat{X}_B^{(k)}(t)\}_{1 \leq k \leq N}$ using  $Z_B(t)$, where $\hat{X}_B^{(k)}(t)$ denotes the estimate at node~$k$ at time $t$ if $B(t)=B$. 
Update $f^{(t)}(B)$ using \eqref{eqn:fB-update-iid-data-distributed}, and $K_{B}^{(t)}$ using \eqref{eqn:coefficients-update-iid-data-distributed} for all $B \in \mathcal{B}$.

3. Do the same $\theta(t)$ update as in \eqref{eqn:theta-update-iid-data}.

4. Broadcast $K_{B(t)}^{(t+1)}$, $f^{(t)}(B)$ and $\theta(t+1)$  if $\mathcal{J}(t)=1$.

\label{algorithm:distributed-gibbs-learning-algorithm-for-constrained-problem-EM}
 \end{algorithm}  
 }}

\begin{remark}
 Algorithm~\ref{algorithm:distributed-gibbs-learning-algorithm-for-constrained-problem-EM} is similar to Algorithm~\ref{algorithm:gibbs-learning-algorithm-for-constrained-problem-EM} except that (i) consensus is used for deciding the estimates, and (ii) an additional SPSA algorithm   has been used to optimize the consensus gains. However,   this scheme does not achieve perfect consensus, and is optimal only when one round message exchange among neighbouring nodes is allowed per slot.
\end{remark}
\begin{remark}
Since $K_B^{(t)}$ iteration does not depend on $\lambda(t)$, we can run \eqref{eqn:coefficients-update-iid-data-distributed} at the same timescale as $\lambda(t)$ iteration.
\end{remark}

\subsection{Performance of Algorithm~\ref{algorithm:distributed-gibbs-learning-algorithm-for-constrained-problem-EM}}

\subsubsection{Complexity and distributed nature}\label{subsubsection:complexity-distributed-iid}
{\em The mean number of additional sensors activated per slotis $O(\frac{N}{T})$, which can be made small by taking $T$ large enough. The same argument applies to the computation of $Y_{K_B^{(t)}}$. Moreover, one can decide only to compute $Y_{K_{B(t)}^{(t)}}(t)$ and update $f^{(t)}(B(t))$ when $\mathcal{J}(t)=1$ as discussed in Section~\ref{subsubsection:computational-complexity-of-centralized-iid-tracking}; the algorithm will  converge slowly but $O(2^N)$ computation will be avoided.}

{\em Gibbs sampling is run at all nodes; they will yield the same $B(t)$ since all sensors have the same seed. All sensors will have the same $\lambda(t)$,   and  can update $h^{(t)}(B(t))$ for each $t$.  The consensus gains and $f^{(t)}(B)$ updates need to be sent to all sensors if $\mathcal{J}(t)=1$; however, a bounded delay in broadcast  does not affect convergence. {\bf Since nodes use only local consensus and periodic broadcast, and Gibbs sampling step is distributed, Algorithm~\ref{algorithm:distributed-gibbs-learning-algorithm-for-constrained-problem-EM} is distributed.}}

\subsubsection{Convergence of Algorithm~\ref{algorithm:distributed-gibbs-learning-algorithm-for-constrained-problem-EM}}
\begin{assumption}\label{assumption:existence-of-optimal-linear-combination}
For given  $\lambda \in [0,A_0]$, $\theta \in \Theta$, for all $B \in \mathcal{B}$, the function $\mathbb{E}(Y_{K_B})$ of $K_B$ (Section~\ref{subsubsection:Y_K_B-and-f-t-B-update})  is Lipschitz continuous in $K_B$. The set of ordinary differential equations $\dot{K}_B(\tau)=-\nabla \mathbb{E}(Y_{K_B(\tau)}(\tau))$ (vectorized)  for any fixed $\theta \in \Theta$ has a globally asymptotically stable equilibrium $K_B^*(\theta)$ which is Lipschitz continuous in $\theta$.
\end{assumption}

Now we present the main result related to Algorithm~\ref{algorithm:distributed-gibbs-learning-algorithm-for-constrained-problem-EM}, which shows that the iterates of Algorithm~\ref{algorithm:distributed-gibbs-learning-algorithm-for-constrained-problem-EM} almost surely converge to a set of locally optimal solutions of \eqref{eqn:distributed-constrained-problem}.

\noindent\fbox{
    \parbox{0.46 \textwidth}{
\begin{theorem}\label{theorem:convergence-of-distributed-estimation-iid-data}
Under Assumptions~\ref{assumption:Lipschitz-continuity-wrt-theta}-\ref{assumption:existence-of-optimal-linear-combination} and Algorithm~\ref{algorithm:distributed-gibbs-learning-algorithm-for-constrained-problem-EM}, we have  $\lim_{t \rightarrow \infty}\theta(t)=\theta^*$ almost surely. Correspondingly, $\lambda(t) \rightarrow \lambda^*(\theta^*)$ almost surely. 
As a result, for all $B \in \mathcal{B}$, we have $K_B(t) \rightarrow K_B^*(\theta^*)$  almost surely. 
Also, $f^{(t)}(B) \rightarrow \frac{1}{N} \sum_{k=1}^N \mathbb{E}_{Z_B(t) \sim p(\cdot| \theta_0)} \mathbb{E}_{\mu_2^{(k)}(\cdot;\cdot;\theta^*), K_B^*(\theta^*),  \theta^*} (||X(\tau)-\hat{X}^{(k)}(\tau)||^2|Z_B(t)):=f_{\theta^*}(B)$ almost surely for all $B \in \mathcal{B}$.  The $B(t)$ process reaches the steady-state distribution  $\pi_{\beta,\lambda^*(\theta^*), K_B^*(\theta^*), \theta^*}(\cdot)$ which can be obtained by replacing $h(B)$ in \eqref{eqn:definition-of-Gibbs-distribution} by $f_{\theta^*}(B)+\lambda^*||B||_1$.
\end{theorem}
}}
\begin{proof}
The proof is similar to that of Theorem~\ref{theorem:convergence-of-GIBBSLEARNINGEM}. 
\end{proof}

If $\theta^*=\theta_0$, then  global optimum for \eqref{eqn:distributed-constrained-problem} is reached.

\section{Distributed tracking of a Markov chain}
\label{section:markovian-data}
In this section, we seek to track a Markov chain $\{X(t)\}_{t \geq 0}$ with transition probability matrix $A^{T}$ and finite state space $\mathcal{X}$. In order to have a meaningful problem, we enumerate each state and denote the $i$-th state by $\mathbf{e}_i$ which is an $|\mathcal{X}|$-length $0-1$ column vector with $1$ only at the $i$-th location. Thus, the state space becomes $\mathcal{X}=\{\mathbf{e}_i: 1 \leq i \leq |\mathcal{X}|\}$. We also consider a measurement process where, given $X(t)=\mathbf{e}_i$ and a configuration $B(t)=B$ of active sensors, any sensor~$k$ with $B_k=1$ makes an observation $z_k(t) \sim \mathcal{N}(m_{k,i}, \Sigma_{k,i})$ where the mean $m_{k,i}$ and covariance matrix $\Sigma_{k,i}$ depend on the state $\mathbf{e}_i$. Under this model, a centralized, finite horizon  version of the dynamic sensor subset selection problem has been solved in \cite{daphney-etal14active-classification-pomdp}, where it is shown that a sufficient statistic for decision making is a belief vector on the state space conditioned on the history. The authors of \cite{daphney-etal14active-classification-pomdp} formulated a partially observable Markov decision process (POMDP) with this belief vector working as a proxy to the (hidden) state, and also proposed a Kalman-like estimator to make on-line update to the belief vector using new observations made by chosen sensors.  Hence, in this section, we skip the centralized problem~\eqref{eqn:centralized-constrained-problem} and directly solve the distributed   problem~\eqref{eqn:distributed-constrained-problem}.  

The centralized  problem in \cite{daphney-etal14active-classification-pomdp} itself leads to intractability in the  sequential subset selection problem; the POMDP formulation in \cite{daphney-etal14active-classification-pomdp} does not provide any structural result on the optimal policy. Hence, for solving the distributed problem~\eqref{eqn:distributed-constrained-problem}, we restrict ourselves to the class of {\em myopic} policies which seek to minimize the cost at the current time instant.

The estimation scheme yields $\hat{X}^{(k)}(t)$ at each node~$k$; $\hat{X}^{(k)}(t)$ is the belief vector on the state space, and is generated by the use of a Kalman-consensus filter (KCF) as in \cite{saber09kalman-consensus-optimality-stability}. Consensus is required since all nodes do not have access to the complete observation set; consensus requires each sensor to combine the estimates made by its neighbouring sensors. Kalman filtering operation is required since the dynamical system can be expressed as a linear stochastic system  (for which Kalman filter is the best linear MMSE estimator):

\footnotesize
\begin{eqnarray*}
X(t+1)&=&A X(t)+ \underbrace{(X(t+1)-A X(t))}_{:=w(t)}\\
z_k(t+1)&=&   \underbrace{[ m_{k,1}, m_{k,2}, \cdots, m_{k,|\mathcal{X}|}]}_{:=H_k} X(t)+ \underbrace{v_k(t)}_{\sim \mathcal{N}(m_{k,i},\Sigma_{k,i}) \text{ if } X(t)=\mathbf{e}_i} \\
Z_k(t)&=&B_k(t) z_k(t)
\end{eqnarray*}
\normalsize
Since $A$ is known,  the conditional covariance matrix $Q^{(i)}$ of $w(t)$ given  $X(t)=\mathbf{e}_i$ is known to all sensors. 

\subsection{Kalman consensus filtering}\label{subsection:KCF-equations-from-Saber}
 The KCF that we use is adapted from \cite{saber09kalman-consensus-optimality-stability} with the additional consideration of $B(t)$. Here each sensor~$k$ maintains an estimate $\bar{X}^{(k)}(t)$ at time $t$ before making any observation. Once the observations are made, sensor~$k$ computes  $\hat{X}^{(k)}(t)$ by using KCF. The estimates evolve  as follows:

\footnotesize
\begin{eqnarray}
\hat{X}^{(k)}(t)&=&[\bar{X}(t)+K_k(t) B_k(t) (z_k(t)-H_k \bar{X}^{(k)}(t)) \nonumber\\
&& + C_k \sum_{j \in nbr(k)} (\bar{X}^{(j)}(t)-\bar{X}^{(k)}(t))]_{\mathcal{D}} \nonumber \\
\bar{X}^{(k)}(t+1)&=& A \hat{X}^{(k)}(t) \label{eqn:state-update-KCF}
\end{eqnarray}
\normalsize
Here $C_k$ and $K_k(t)$ are called {\em consensus gain} and {\em Kalman gain} matrices for sensor~$k$, and $nbr(k)$ is the set of neighbours of sensor~$k$. Projection on the probability simple $\mathcal{D}$ is done to ensure  that $\hat{X}^{(k)}(t)$ is a valid probability belief vector.

In \cite[Theorem~$1$]{saber09kalman-consensus-optimality-stability},  the Kalman  gains $K_k(t)$ at all nodes are optimized for given consensus gains $C_k$, so that the MSE $\frac{1}{N}\sum_{k=1}^N \mathbb{E}||\hat{X}^{(k)}(t)-X(t)||^2$ at the current time step is minimized; but the computational and communication complexity per node for its implementation grows rapidly with $N$. Hence,  \cite[Section~IV]{saber09kalman-consensus-optimality-stability} also provides an alternative {\em suboptimal} KCF algorithm to compute $K_k(t)$ at each sensor~$k$, which has low complexity and is easily implementable.  Hence, we adapt  this   suboptimal algorithm from  \cite[Section~IV]{saber09kalman-consensus-optimality-stability}  to our problem. 

The KCF gain update scheme from \cite[Section~IV]{saber09kalman-consensus-optimality-stability} maintains two matrices $M_k(t)$ and $P_k(t)$ for each $k$, which are viewed as proxies for the  covariance matrices of the two errors $\hat{X}^{(k)}(t)-X(t)$ and $\bar{X}^{(k)}(t)-X(t)$. It also requires system noise covariance matrix; since system noise $w(t)$ is dependent on $X(t)$ whose exact value is unknown to the sensors, we use $Q_k(t)=\sum_{i=1}^N \bar{X}^{(k)}(t+1)(i) Q^{(i)} $ at node~$k$ as an estimate of the covariance matrix of $w(t+1)$. Similarly, 
$R_k(t+1)= B_k(t)  \sum_{i=1}^{|\mathcal{X}|} \bar{X}^{(k)}(t+1)(i) \Sigma_{k,i}$ is used as an alternative to the covariance matrix $v_k(t+1)$. The KCF filter also maintains an abstract iterate $F_k(t)$. The overall KCF gain update equations from \cite[Section~IV]{saber09kalman-consensus-optimality-stability} are as follows:

\footnotesize
\begin{eqnarray}
K_k(t)&=&B_k(t) P_k(t) H_k^T (R_k(t)+ H_k P_k(t)H_k^T)^{-1} \nonumber\\
 R_k(t+1)&=& B_k(t)  \sum_{i=1}^{|\mathcal{X}|} \bar{X}^{(k)}(t+1)(i) \Sigma_{k,i}  \nonumber\\
 Q_k(t) &=& \sum_{i=1}^N \bar{X}^{(k)}(t+1)(i) Q^{(i)} \nonumber\\
F_k(t) &=& I-K_k (t) H_k B_k(t) \nonumber\\
M_k(t) &=& F_k(t) P_k(t) F_k^T(t)+K_k(t) R_k(t+1) K_k^T(t)\nonumber\\
P_k(t+1) &=& AM_k(t) A^T+ Q_k(t) \label{eqn:KCF-gain-update-equation}
\end{eqnarray}
\normalsize

\subsection{The proposed algorithm}
The sensor subset selection is done via Gibbs sampling run at all nodes  supplied with the same seed;  all nodes generate the same configuration $B(t)$ at each time $t$. The quantity $f^{(t)}(B)$ is updated via stochastic approximation to converge to the MSE under configuration $B$; but since the MSE under $B$ cannot be computed directly, sensor $k$ uses $\frac{1}{N} \sum_{k=1}^N Tr(M_k(t-T))$ of a past slot $(t-T)$ to update $f^{(t)}(B)$. $\lambda(t)$  is varied at a slower timescale.

The proposed scheme   is provided in Algorithm~\ref{algorithm:distributed-tracking-of-a-Markov-chain}. 

\noindent\fbox{
    \parbox{0.46 \textwidth}{
\begin{algorithm}

{\bf Input:}  $A_0$, $T>0$  stepsize sequences $\{a(t)\}_{t \geq 0}$ and $\{b(t)\}_{t \geq 0}$ as in Section~\ref{section:iid-data}, consensus gain matrices $C_k$ for all $k \in \{1,2,\cdots,N\}$, the same seed for randomization to all sensors, initial covariance matrix $P(0)$ of $X(0)$. All iterates are initialized arbitrarily. Define $\nu_B(t):=\sum_{\tau=0}^t \mathbb{I}(B(\tau)=B) \mathcal{J}(t)$.

{\bf For any time $t=0,1,2,\cdots$, do at each sensor~$k$:}

1.  Select  $B(t)$ at each sensor by running the Gibbs sampling step as in Section~\ref{subsubsection:gibbs-sampling-step-iid-centralized}. Activate sensors autonomously according to the common $B(t)$ selected by them, and make observations accordingly.

2. At  sensor~$k$, perform  state estimation with \eqref{eqn:state-update-KCF} and gain update with  \eqref{eqn:KCF-gain-update-equation}. Update $\lambda(t)$ at all sensors using \eqref{eqn:lambda-update-iid-data}. Compute 
$h^{(t+1)}(B) = f^{(t+1)}(B)+\lambda(t) ||B||_1$ either for all $B \in \mathcal{B}$, or for $B(t)$ if computation is a bottleneck.

3. If $\mathcal{J}(t)=1$, update:

\footnotesize
\begin{eqnarray*}
 f^{(t+1)}(B)&=&[ f^{(t)}(B)+\mathcal{J}(t) \mathbb{I}(B(t-T)=B)  a(\nu_B(t-T))   \\
&&  (\frac{1}{N}\sum_{k=1}^N Tr(M_k(t-T)) - f^{(t)}(B)) ]_0^{A_0} 
\end{eqnarray*}
\normalsize

4. If $\mathcal{J}(t)=1$, broadcast $Tr(M_k(t))$ to all sensors, so that step~$3$ can be performed at time $(t+T)$.

\label{algorithm:distributed-tracking-of-a-Markov-chain}
 \end{algorithm}  
 }}

\begin{remark}
Algorithm~\ref{algorithm:distributed-tracking-of-a-Markov-chain} is suboptimal for \eqref{eqn:distributed-constrained-problem} because: (i) it greedily chooses $B(t)$ via Gibbs sampling  without caring about the future cost, and  (ii) KCF update is suboptimal. {\em But it has low  complexity, and it performs well   (see Section~\ref{section:numerical-results}).}
\end{remark}

\subsection{Complexity of Algorithm~\ref{algorithm:distributed-tracking-of-a-Markov-chain}}
 {\em At each time $t$, a sensor~$k$ needs to obtain $\bar{X}^{(j)}(t)$ from all its neighbours $j \in nbr(k)$ for consensus. Also, $\{Tr(M_k(t))\}_{1 \leq k \leq N}$ needs to be broadcast to all nodes when $\mathcal{J}(t)=1$ so that $f^{(t+T)}(B)$ update can be done at each node; but the per slot communication for this broadcast can be made small enough by making  $T$   arbitrarily large but finite, and the broadcast can even be done over $T$ slots to avoid network congestion at any particular slot. Interestingly,   $M_k(t)$ and $P_k(t)$  can be updated using local iterates, and do not need any communication. Computing $M_k(t)$ and $P_k(t)$ involves simple matrix operations which have polynomial complexity.}

\section{Numerical Results}\label{section:numerical-results}
Now we   demonstrate the performance of Algorithm~\ref{algorithm:gibbs-learning-algorithm-for-constrained-problem-EM} (centralized) and Algorithm~\ref{algorithm:distributed-tracking-of-a-Markov-chain} (distributed). We consider the following parameter values: $N=5$, $\bar{N}=2$, $a(t)=\frac{1}{t^{0.6}}$, $b(t)=\frac{2}{t^{0.8}}$, $c(t)=\frac{1}{t}$, $d(t)=\frac{0.2}{t^{0.1}}$, $T=20$, $\lambda(0)=0.1$, $\beta=150$. Gibbs sampling is run $10$ times per slot.

\subsubsection{Performance of Algorithm~\ref{algorithm:gibbs-learning-algorithm-for-constrained-problem-EM}}
For illustration purpose, we assume that $X(t)  \sim \mathcal{N}(\theta_0, (1-\theta_0)^2)$ {\em scalar}, and $z_k(t)=X(t)+w_k(t)$, where $\theta_0=0.5$  and $w_k(t)$ is zero mean  i.i.d. Gaussian noise independent across $k$. Standard deviation of $w_k(t)$ is chosen uniformly and independently from the interval $[0,0.5]$, for each $k \in \{1,2,\cdots,N\}$.  Initial estimate $\theta(0)=0.2$, $\Theta=[0,0.8]$. 

\begin{figure*}[t]
\begin{minipage}[c]{0.5\linewidth}
\subfigure{
\includegraphics[height=6cm, width=\linewidth]{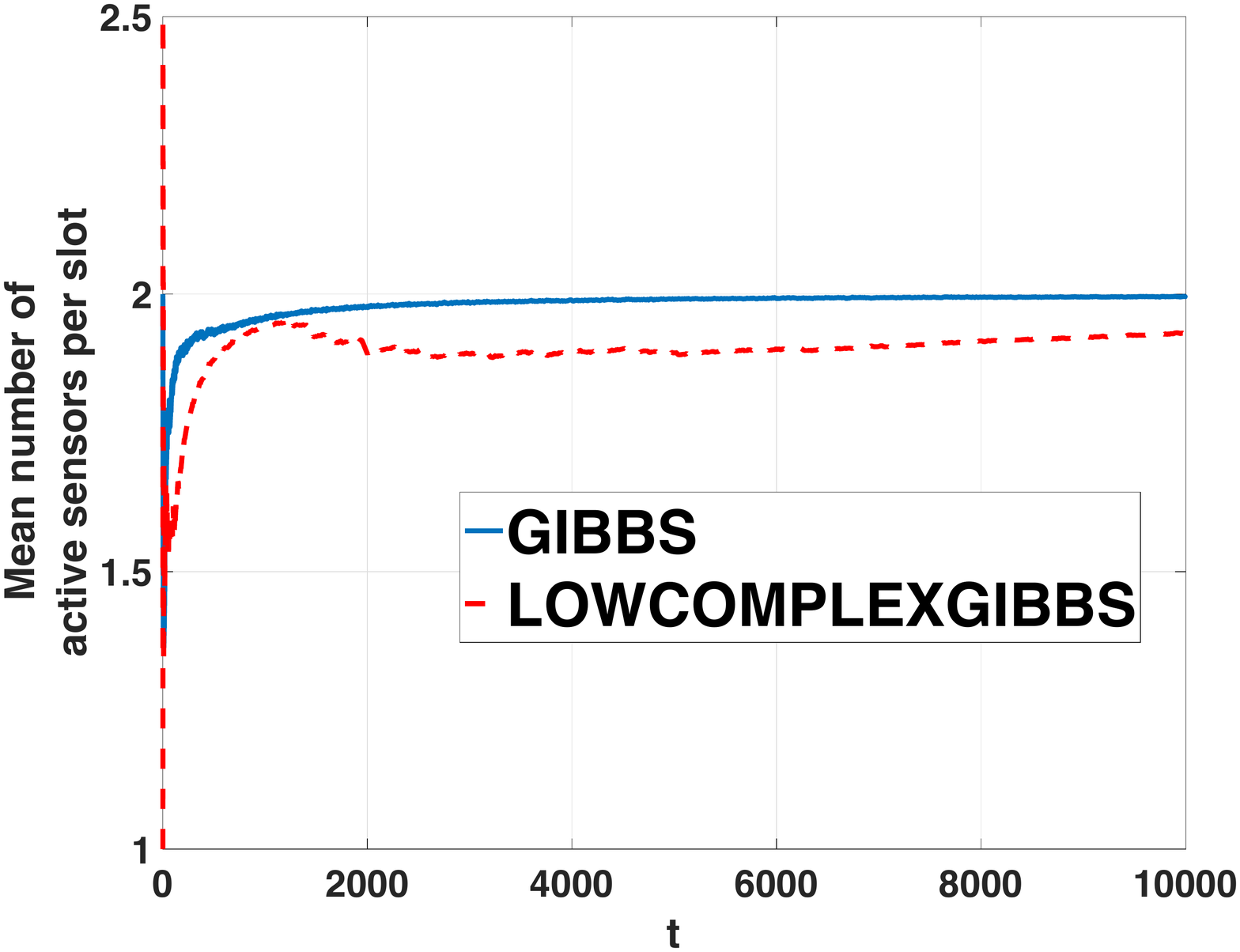}
\includegraphics[height=6cm, width=\linewidth]{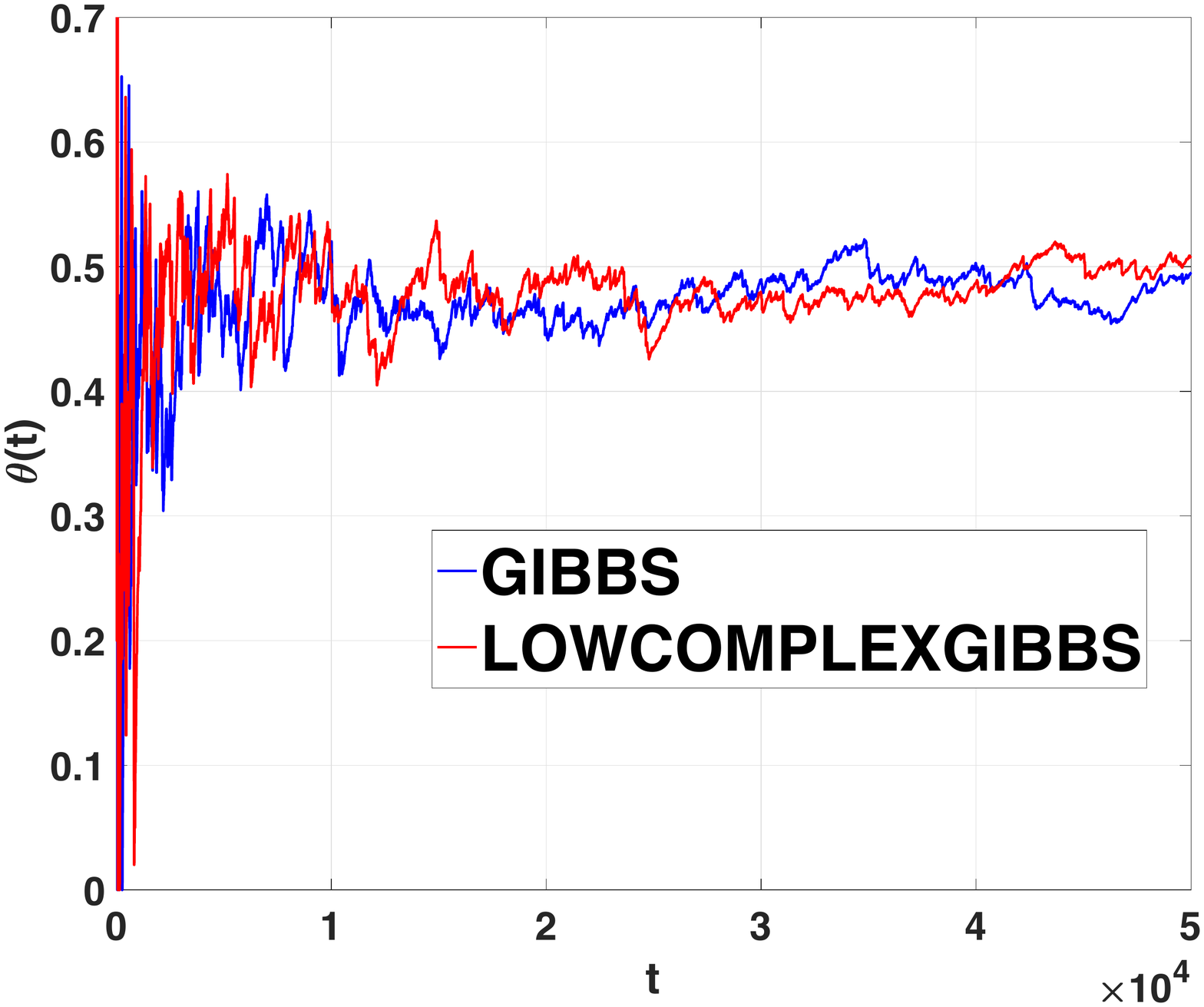}}
\end{minipage} \hfill
\caption{Performance of Algorithm~\ref{algorithm:gibbs-learning-algorithm-for-constrained-problem-EM} for centralized tracking of the iid process.}
\label{fig:performance-of-centralized-iid-tracking}
\end{figure*}

We consider three possible algorithms: (i) Algorithm~\ref{algorithm:gibbs-learning-algorithm-for-constrained-problem-EM} in its basic form, which we call GIBBS, (ii) a variation of Algorithm~\ref{algorithm:gibbs-learning-algorithm-for-constrained-problem-EM} called LOWCOMPLEXGIBBS where all sensors are not read when $\mathcal{J}(t)=1$, and the relatively expensive $f^{(t)}(B(t))$ and $\theta(t)$ updates are done  every $T$ slots, and (iii) an algorithm  GREEDY where   $\bar{N}$ sensors are picked arbitrarily and used for ever with the wrong estimate $\theta(0)=0.2$ without any update. 
The  MSE per slot, mean number of active sensors per slot, $\lambda(t)$ and $\theta(t)$ are plotted against   $t$ in Figure~\ref{fig:performance-of-centralized-iid-tracking}.  MSE  of all these three algorithms are much smaller than  $Var(X(t))=(1-\theta_0)^2$ (this is MMSE  without any observation). We notice that GIBBS and LOWCOMPLEXGIBBS significantly outperform GREEDY in terms of time-average MSE; this shows the power of Gibbs sampling and learning $\theta(t)$ over time.  We have plotted only one sample path since  GIBBS and LOWCOMPLEXGIBBS  converge almost surely to the global optimum. We also observe that GIBBS converges faster than LOWCOMPLEXGIBBS, since it uses more computational and communication resources. We observe that 
$\frac{1}{t}\sum_{\tau=1}^t ||B(\tau)||_1 \rightarrow \overline{N}$ and $\theta(t) \rightarrow \theta_0$ almost surely 
for both GIBBS and LOWCOMPLEXGIBBS (verified by  simulating  multiple sample paths). It is interesting to note that $\theta^*=\theta_1=\theta_0$ in this numerical example (recall Theorem~\ref{theorem:convergence-of-GIBBSLEARNINGEM} and Remark~\ref{remark:what-if-all-sensors-are-not-read}), i.e., both GIBBS and LOWCOMPLEXGIBBS converge to the true parameter. Convergence rate   will vary with stepsize   and other parameters.

 \begin{figure*}[t]
 \begin{minipage}[r]{0.5\linewidth}
\subfigure{
\includegraphics[height=6cm, width=\linewidth]{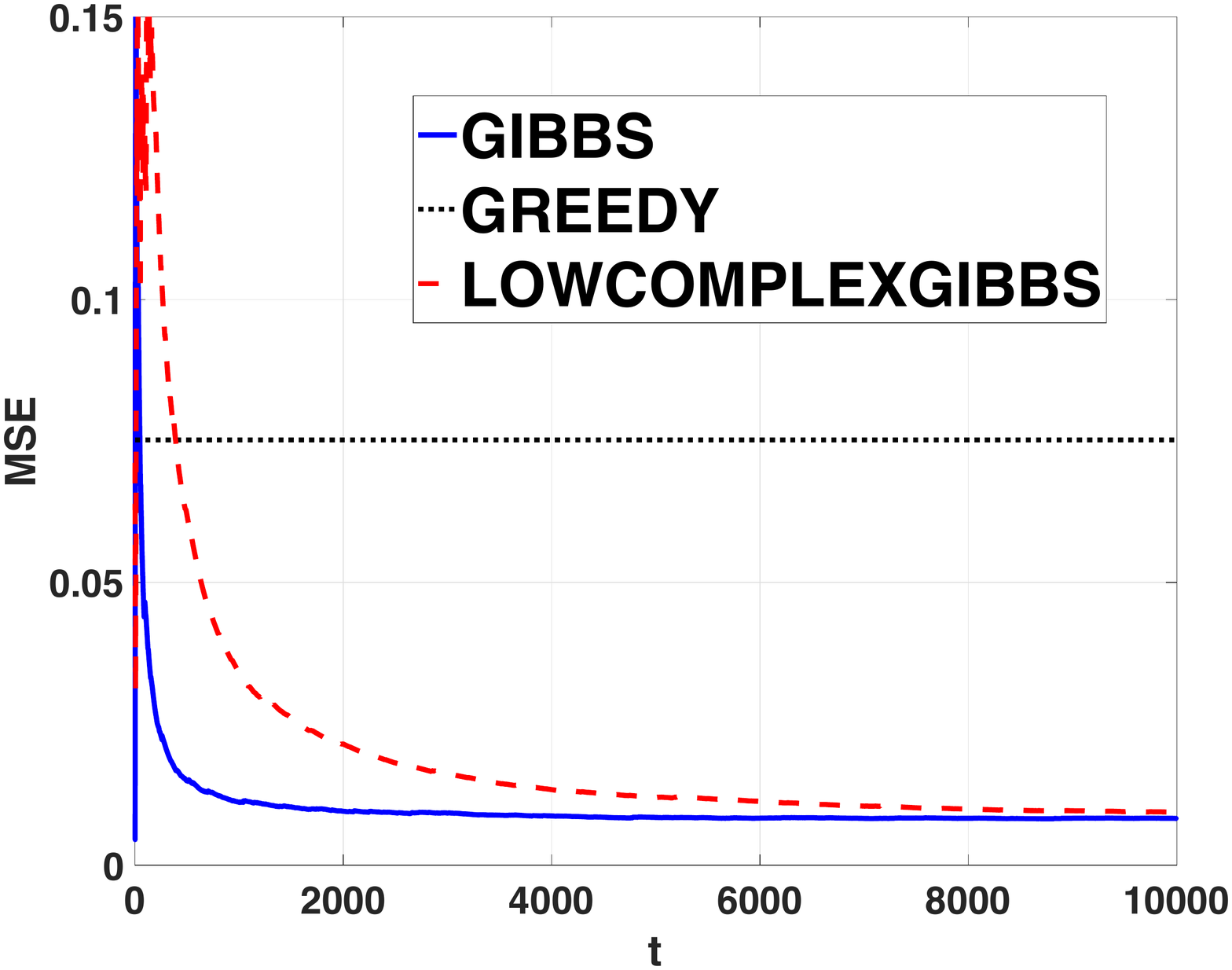}
\includegraphics[height=6cm, width=\linewidth]{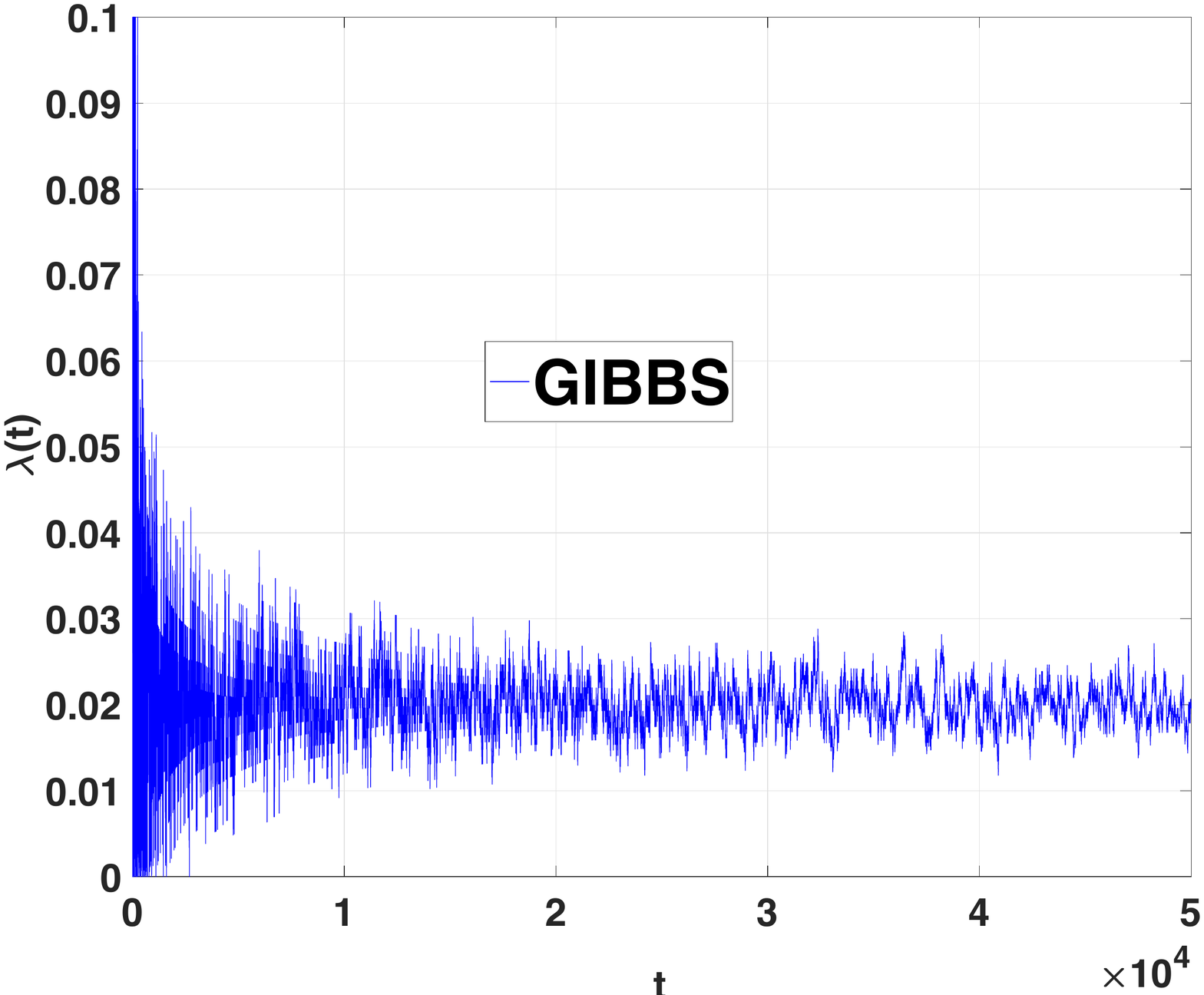}}
\end{minipage}  \hfill
\end{figure*}

\subsubsection{Performance of Algorithm~\ref{algorithm:distributed-tracking-of-a-Markov-chain}}
We consider $|\mathcal{X}|=4$ states, and assume that the sensors form a line topology. 
Transition probability matrix $A^T$ is chosen randomly, and $C_k=0.1$ is set in Algorithm~\ref{algorithm:distributed-tracking-of-a-Markov-chain} (which we call GIBBSKCF here) for all $k$. The values $m_{k,i}$ are chosen uniformly from $[0,1]$, and $\Sigma_{k,i}=0.05*(1+|k-i|)$ (scalar values in this case) are set for all  $1 \leq i \leq |\mathcal{X}|, 1 \leq k \leq N$. 

We compared the MSE performance of three algorithms: (i) GIBBSKCF, i.e., Algorithm~\ref{algorithm:distributed-tracking-of-a-Markov-chain}, (ii) CENTRALKALMAN, where a centralized Kalman filter tracks $X(t)$ using observations only from two arbitrary sensors, and (iii) PERFECTBLIND, where at each time $t$, the state $X(t-1)$ is known perfectly to all sensors, but no observation is allowed to compute $\hat{X}^{(k)}(t)$. The MSE of PERFECTBLIND will be $\lim_{t \rightarrow \infty}\mathbb{E}(Var(X(t)|X(t-1)))$. 

\begin{figure}[t!]
\begin{centering}
\begin{center}
\includegraphics[height=6cm, width=7.5cm]{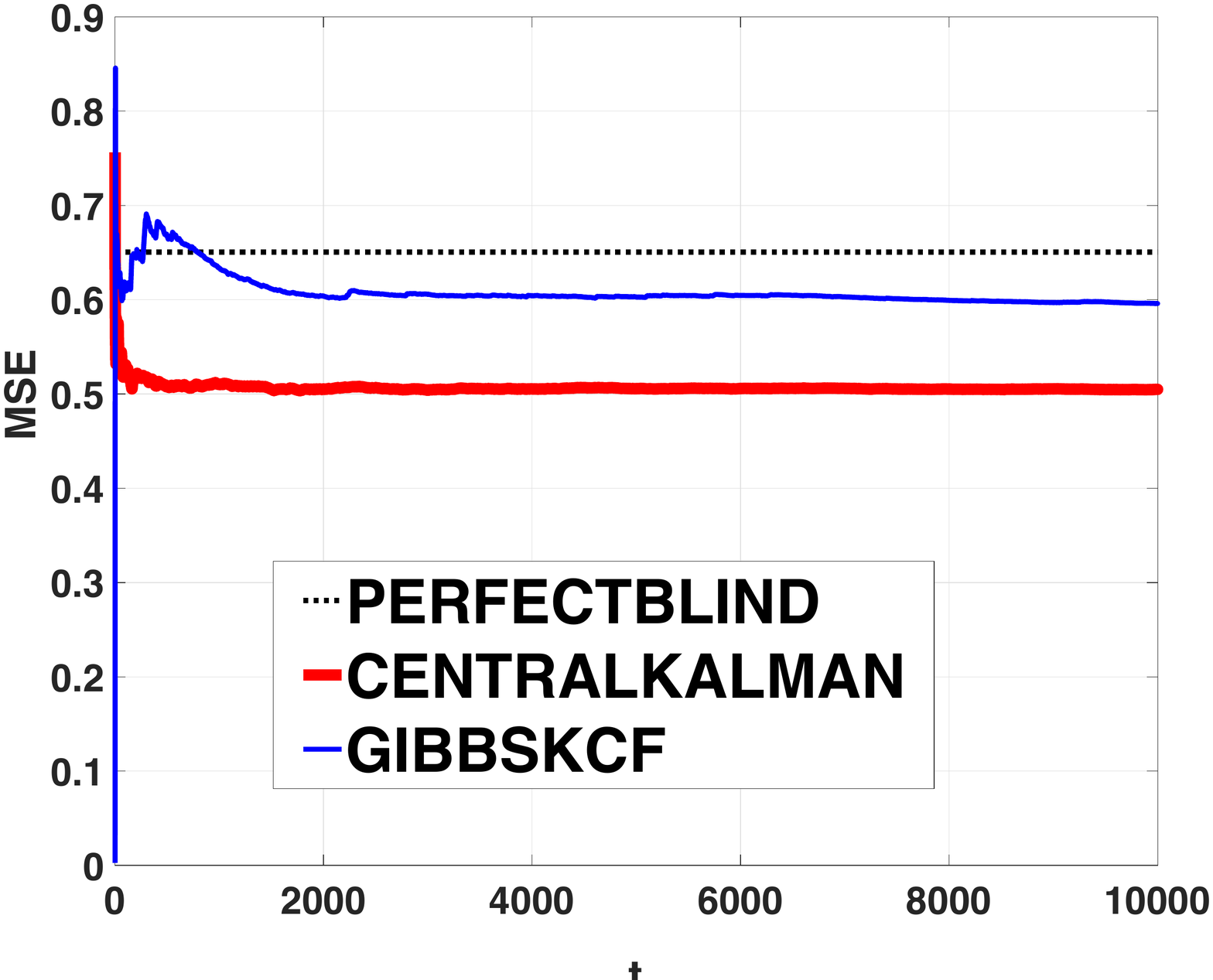}
\includegraphics[height=6cm, width=7.5cm]{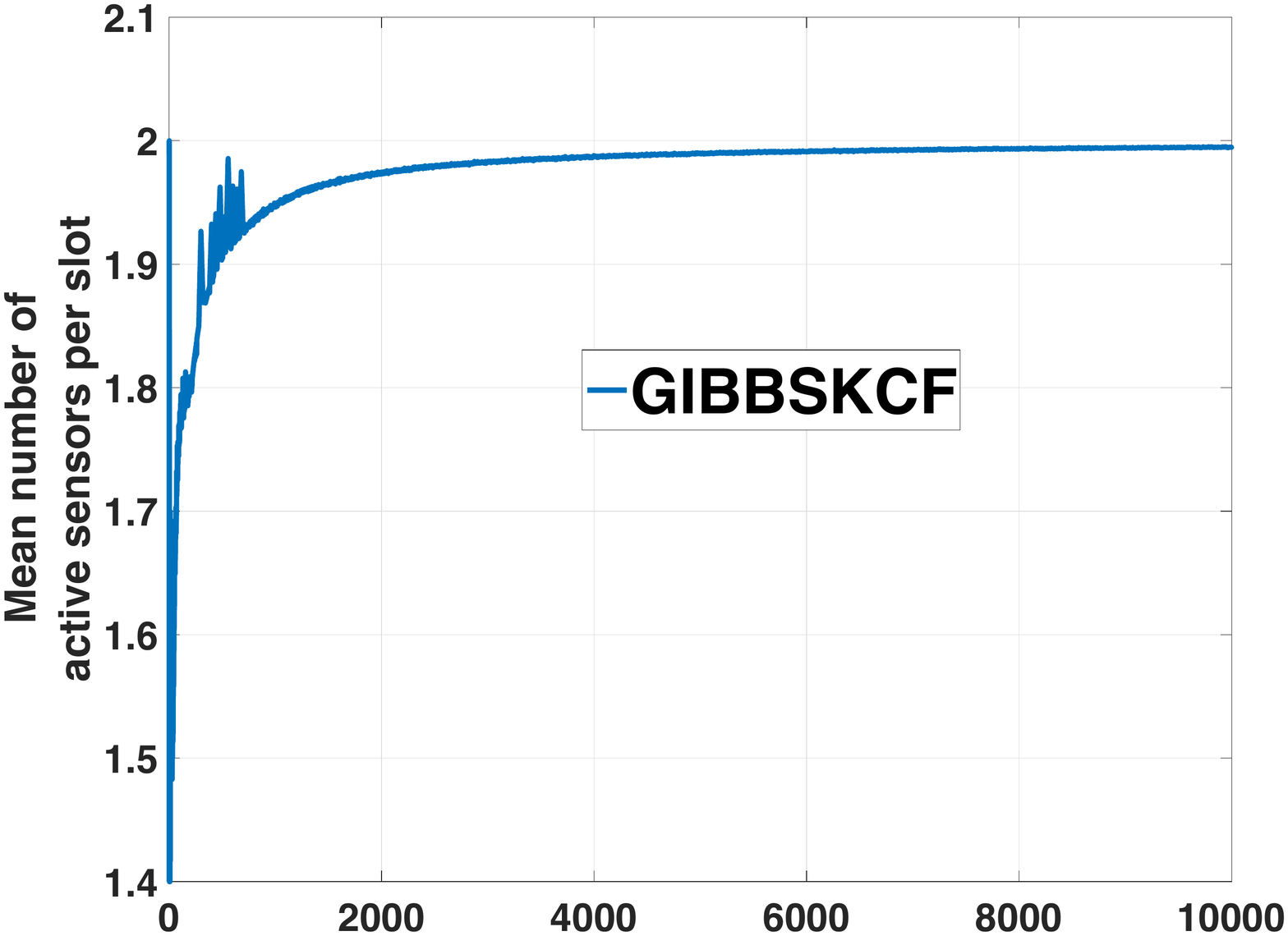}
\end{center}
\end{centering}

\caption{Performance of Algorithm~\ref{algorithm:distributed-tracking-of-a-Markov-chain} for distributed tracking of a Markov chain.}

\label{fig:performance-of-distributed-tracking-of-a-Markov-chain}
\end{figure}

In Figure~\ref{fig:performance-of-distributed-tracking-of-a-Markov-chain}, we plot the time-average MSE for all three algorithms along one sample path,  and $\frac{1}{t}\sum_{\tau=1}^t ||B(\tau)||_1$ for GIBBSKCF.  We observe that, $\frac{1}{t}\sum_{\tau=1}^t ||B(\tau)||_1$ converges to $\bar{N}$.  For the given sample path, GIBBSKCF provides better MSE than PERFECTBLIND, but the MSE was seen to be slightly worse along some other sample paths. We also observe that the MSE of GIBBSKCF is {\em slightly} worse than CENTRALKALMAN for the given instance, but they are very close in many other problem instances (verified numerically); this establishes the efficacy of  GIBBSKCF  and the power of Gibbs sampling based sensor subset selection, despite using only one round of consensus per slot. Basically, dynamic subset selection compensates for the performance loss due to lack of a fusion center.

\section{Conclusions}\label{section:conclusion}

We have proposed low-complexity centralized and distributed learning algorithms for dynamic sensor subset selection for tracking i.i.d. time-varying as well as Markovian processes. We first provided  algorithms based on Gibbs sampling and stochastic approximation for i.i.d. time-varying data with unknown parametric distribution, and proved almost sure convergence. Next, we provided an algorthm based on Kalman consensus filtering, Gibbs sampling and stochastic approximation for distributed tracking of a  Markov chain.  Numerical results demonstrate the efficacy of the algorithms against simple algorithms without learning.

{\small
\bibliographystyle{unsrt}
\bibliography{arpan-techreport}
}

\vspace{-20mm}

\begin{IEEEbiography}[{\includegraphics[width=1in,height=1in,clip,keepaspectratio]{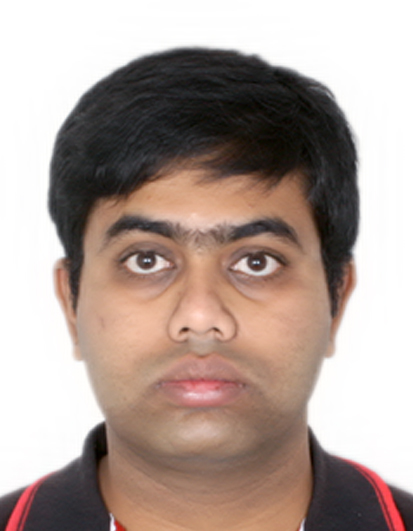}}]{Arpan 
Chattopadhyay} obtained his B.E. in Electronics and Telecommunication Engineering from Jadavpur University, 
Kolkata, India in the year 2008, and M.E. and Ph.D in Telecommunication Engineering from Indian Institute of Science, 
Bangalore, India in the year 2010 and 2015, respectively. He is currently working in the Ming Hsieh Department of Electrical Engineering, University of Southern California, Los Angeles  as a postdoctoral researcher. Previously he worked as a postdoc in INRIA/ENS Paris. 
His research interests include wireless networks, cyber-physical systems, machine learning and control. \end{IEEEbiography}

\begin{IEEEbiography}{Urbashi Mitra} is a Dean’s Professor in the Departments of Electrical
Engineering and Computer Science, University of Southern California, Los Angeles,
CA, USA. Previous appointments include Bellcore and the Ohio State
University. Her honors include: U.S.-U.K. Fulbright, Leverhulme Trust Visiting
Professorship, Editor-in-Chief IEEE TRANSACTIONS ON MOLECULAR, BIOLOGICAL
AND MULTISCALE COMMUNICATIONS, IEEE Communications Society
Distinguished Lecturer, U.S. NAE Galbreth Lectureship, Okawa Foundation
Award, and an NSF CAREER Award. Her research is in wireless
communications. \end{IEEEbiography}

\newpage

\appendices

\section{Proof of Theorem~\ref{theorem:relation-between-constrained-and-unconstrained-problems}}
\label{appendix:proof-of-relation-between-constrained-and-unconstrained-problems}
We will prove only the first part of the theorem where there exists one $B^*$. The second part of the theorem can be proved similarly. 

Let us denote the optimizer for \eqref{eqn:centralized-constrained-problem} by $B$, which is possibly different from $B^*$. Then, by the definition of $B^*$, we have $f(B^*)+ \lambda^* ||B^*||_1 \leq f(B)+ \lambda^* ||B||_1$. But $||B||_1 \leq K$ (since $B$ is a feasible solution to the constrained problem) and $||B^*||_1=K$ (by assumption). Hence, $f(B^*) \leq f(B)$. This completes the proof.

\section{Weak and Strong Ergodicity}
\label{appendix:weak-and-strong-ergodicity}
Consider a discrete-time  Markov chain (possibly not time-homogeneous) $\{B(t)\}_{t \geq 0}$ with transition probability matrix (t.p.m.)  $P(m;n)$ between 
$t=m$ and $t=n$. We denote by $\mathcal{D}$   the collection of all possible probasbility distributions 
 on the state space. Let $d_V(\cdot,\cdot)$ denote  the total variation distance between two distributions in $\mathcal{D}$. 
Then $\{B(t)\}_{t \geq 0}$ is called weakly ergodic if, for all $m \geq 0$,  we have 
$\lim_{n \uparrow \infty} \sup_{\mu,\nu \in \mathcal{D}}  d_V (\mu P(m;n) , \nu P(m;n) ) =0 $.

The Markov chain $\{B(t)\}_{t \geq 0}$ is called strongly ergodic if there exists $\pi \in \mathcal{D}$ such that, 
$\lim_{n \uparrow \infty} \sup_{\mu \in \mathcal{D}}  d_V (\mu^{T} P(m;n) , \pi ) =0 $ for all $m \geq 0$.

\section{Proof of Theorem~\ref{theorem:result-on-weak-and-strong-ergodicity}}
\label{appendix:proof-of-weak-and-strong-ergodicity}
 
We will first show that the Markov chain $\{B(t)\}_{t \geq 0}$ in weakly ergodic.

Let us define $\Delta:=\max_{B \in \mathcal{B}, A \in \mathcal{B}}|h(B)-h(A)|$.

Consider the transition probability matrix (t.p.m.) $P_l$ for the inhomogeneous Markov 
chain $\{Y(l)\}_{l \geq 0}$ (where $Y(l):=B(lN)$). The Dobrushin's ergodic coefficient $\delta(P_l)$ is given by 
(see \cite[Chapter~$6$, Section~$7$]{breamud99gibbs-sampling} for definition) 
$\delta(P_l)=1- \inf_{B^{'},B^{''} \in \mathcal{B}} \sum_{B \in \mathcal{B}} \min \{P_l(B^{'},B),P_l(B^{''},B) \}$. 
A sufficient condition for the Markov chain $\{B(t)\}_{t \geq 0}$  
to be weakly ergodic is $\sum_{l=1}^{\infty}(1-\delta(P_l))=\infty$ (by 
\cite[Chapter~$6$, Theorem~$8.2$]{breamud99gibbs-sampling}).

Now, with positive probability, activation states for all nodes are updated over a period of $N$ slots. Hence, $P_l(B^{'},B)>0$ for all $B^{'},B \in \mathcal{B}$. Also, once a node $j_t$ for $t=lN+k$ 
is chosen  in MODIFIEDGIBBS~algorithm, the sampling probability for any activation state in a slot 
is greater than  $\frac{e^{-\beta(lN+k) \Delta}}{2}$. 
Hence, for independent sampling over $N$ slots, we  have, for all pairs $B^{'},B$: 
$$P_l(B^{'},B) >  \prod_{k=0}^{N-1}\bigg( \frac{e^{-\beta(lN+k) \Delta}}{2N} \bigg) >0$$ 
Hence, 
\begin{eqnarray}
&&\sum_{l=0}^{\infty}(1-\delta(P_l)) \nonumber\\
&=& \sum_{l=0}^{\infty} \inf_{B^{'},B^{''} \in \mathcal{B}} \sum_{B \in \mathcal{B}} \min \{P_l(B^{'},B),P_l(B^{''},B) \} \nonumber\\
& \geq & \sum_{l=0}^{\infty} 2^N \prod_{k=0}^{N-1} \bigg( \frac{e^{-\beta(0) \log(1+lN+k) \times \Delta}}{2N} \bigg) \nonumber\\
& \geq &   \sum_{l=0}^{\infty}  \prod_{k=0}^{N-1} \bigg( \frac{e^{-\beta(0) \log(1+lN+N) \times \Delta}}{N} \bigg)  \nonumber\\
& = &  \frac{1}{N^N} \sum_{l=1}^{\infty}  \frac{1}{  (1+lN)^{\beta(0) N \Delta}}  \nonumber\\
& \geq &  \frac{1}{ N^{N+1}} \sum_{i=N+1}^{\infty}  \frac{1}{  (1+i)^{\beta(0) N \Delta}}  \nonumber\\
& = & \infty
 \end{eqnarray}
Here the first inequality uses the fact that the cardinality of $\mathcal{B}$ is $2^N$. The second inequality follows from  replacing $k$ by $N$ in the numerator. The third inequality follows from lower-bounding  
$\frac{1}{(1+lN)^{\beta(0) N \Delta}}$ by $\frac{1}{N}\sum_{i=lN}^{lN+N-1}  \frac{1}{  (1+i)^{\beta(0) N \Delta}} $. 
The last equality follows from the fact that $\sum_{i=1}^{\infty} \frac{1}{i^a}$ diverges for $0 <a<1$.

 Hence, the Markov chain  $\{B(t)\}_{t \geq 0}$ is  weakly ergodic.
 
 In order to prove strong ergodicity of 
 $\{B(t)\}_{t \geq 0}$, we invoke \cite[Chapter~$6$, Theorem~$8.3$]{breamud99gibbs-sampling}. 
 We denote the  t.p.m. of $\{B(t)\}_{t \geq 0}$ at a specific time $t=T_0$ 
 by $Q^{(T_0)}$, which is a given specific matrix. If  $\{B(t)\}_{t \geq 0}$   evolves up to infinite time 
 with {\em fixed} t.p.m. $Q^{(T_0)}$, then it will reach the stationary distribution $\pi_{\beta_{T_0}}(B)= \frac{e^{-\beta_{T_0} h(B)}}{Z_{\beta_{T_0}}}$. 
 Hence, we can claim that Condition~$8.9$ of \cite[Chapter~$6$, Theorem~$8.3$]{breamud99gibbs-sampling} is  satisfied. 
 
 Next, we check Condition~$8.10$ of \cite[Chapter~$6$, Theorem~$8.3$]{breamud99gibbs-sampling}. 
 For any $B \in \arg \min_{B^{'} \in \mathcal{B}} h(B^{'})$,  we can argue that $\pi_{\beta_{T_0}}(B)$ increases with $T_0$ for 
 sufficiently large $T_0$; this can be verified by considering the derivative of $\pi_{\beta}(B)$ w.r.t. $\beta$. For $B \notin \arg \min_{B^{'} \in \mathcal{B}} h(B^{'})$,  the probability 
 $\pi_{\beta_{T_0}}(B)$ decreases with $T_0$ for large $T_0$. Now, using the fact that any monotone, bounded sequence converges, we can write $\sum_{T_0=0}^{\infty} \sum_{B \in \mathcal{B}} |\pi_{\beta_{T_0+1}}(B)-\pi_{\beta_{T_0}}(B)| < \infty$. 
 
 Hence, by \cite[Chapter~$6$, Theorem~$8.3$]{breamud99gibbs-sampling}, the Markov chain $\{B(t)\}_{t \geq 0}$ is strongly ergodic.  It  is straightforward to verify the claim regarding the  limiting distribution.

\section{Proof of Lemma~\ref{lemma:active-sensors-decreasing-in-lambda}}
\label{appendix:proof-of-active-sensors-decreasing-in-lambda}
Let $\lambda_1 > \lambda_2 > 0$, and the corresponding optimal error and mean number of active sensors under these multiplier values be $(f_1,n_1)$ and $(f_2,n_2)$, respectively. Then, by definition, $f_1 + \lambda_1 n_1 \leq f_2 + \lambda_1 n_2$ and $f_2 + \lambda_2 n_2 \leq f_1 + \lambda_2 n_1$. Adding these two inequalities, we obtain $\lambda_1 n_1+\lambda_2 n_2 \leq \lambda_1 n_2+\lambda_2 n_1$, i.e., $(\lambda_1-\lambda_2)n_1 \leq (\lambda_1-\lambda_2)n_2$. Since $\lambda_1>\lambda_2$, we obtain $n_1 \leq n_2$. This completes the first part of the proof. The second part of the proof follows using similar arguments.

\section{Proof of Lemma~\ref{lemma:active-sensors-decreasing-in-lambda-under-basic-gibbs-sampling}}
\label{appendix:proof-of-active-sensors-decreasing-in-lambda-under-basic-gibbs-sampling}
Let us denote $\mathbb{E}_{\mu_2}||B||_1=:g(\lambda)=\frac{\sum_{B \in \mathcal{B}} ||B||_1 e^{-\beta h(B)}}{Z_{\beta}}$. 
It is straightforward to see that $\mathbb{E}_{\mu_2}||B||_1$ is continuously differentiable in $\lambda$.  
Let us denote $Z_{\beta}$ by $Z$ for simplicity. 
The derivative of $g(\lambda)$ w.r.t. $\lambda$ is given by:
\tiny
\begin{eqnarray*}
&& g'(\lambda)\\
&=& \frac{  -Z \beta \sum_{B \in \mathcal{B}} ||B||_1^2 e^{-\beta(f(B)+\lambda ||B||_1)} - \sum_{B \in \mathcal{B}} ||B||_1 e^{-\beta(f(B)+\lambda ||B||_1)}  \frac{dZ}{d \lambda}        }{Z^2}
\end{eqnarray*}
\normalsize

Now, it is straightforward to verify that $\frac{dZ}{d \lambda}=-\beta Z g(\lambda)$. Hence, 
\tiny
\begin{eqnarray*}
&& g'(\lambda)\\
&=& \frac{  -Z \beta \sum_{B \in \mathcal{B}} ||B||_1^2 e^{-\beta(f(B)+\lambda ||B||_1)} + \sum_{B \in \mathcal{B}} ||B||_1 e^{-\beta(f(B)+\lambda ||B||_1)} \beta Z g(\lambda)       }{Z^2}
\end{eqnarray*}
\normalsize

Now, $g'(\lambda) \leq 0$ is equivalent to 
$$g(\lambda) \leq \frac{  \sum_{B \in \mathcal{B}} ||B||_1^2 e^{-\beta(f(B)+\lambda ||B||_1)}    }{    \sum_{B \in \mathcal{B}} ||B||_1 e^{-\beta(f(B)+\lambda ||B||_1)}    }$$
Noting that $\mathbb{E}||B||_1=:g(\lambda)$ and dividing the numerator and denominator of R.H.S. by $Z$, 
the condition is reduced to $\mathbb{E}||B||_1 \leq \frac{\mathbb{E}||B||_1^2}{\mathbb{E}||B||_1}$, which is true since 
$\mathbb{E}||B||_1^2 \geq (\mathbb{E}||B||_1)^2$. Hence, $\mathbb{E}||B||_1$ is decreasing in $\lambda$ for any $\beta>0$. Also, it is easy to verify that $|g'(\lambda)| \leq (\beta+1) N^2$. Hence, $g(\lambda)$ is Lipschitz continuous in $\lambda$.

\section{Proof of Theorem~\ref{theorem:optimality-of-the-learning-algorithm-for-constrained-problem}}
\label{appendix:proof-of-optimality-of-the-learning-algorithm-for-constrained-problem}
Let the distribution of $B(t)$ under Algorithm~\ref{algorithm:gibbs-learning-algorithm-for-constrained-problem}  be 
$\pi^{(t)}(\cdot)$. Since $\lim_{t \rightarrow \infty} a(t)=0$, it follows that $\lim_{t \rightarrow \infty} d_V(\pi^{(t-1)}, \pi_{\beta| \lambda(t-1)})=0$ (where $d_V(\cdot,\cdot)$ is the total variation distance), and $\lim_{t \rightarrow \infty} ( \mathbb{E}_{\pi^{(t-1)}}||B||_1-\mathbb{E}_{\pi_{\beta} | \lambda(t-1) }||B||_1 ) :=\lim_{t \rightarrow \infty} e(t)=0$. Now, we can rewrite the $\lambda(t)$ update equation as follows:
\small
\begin{eqnarray}
 \lambda(t+1)=[  \lambda(t)+a(t) (\mathbb{E}_{\pi_{\beta}|\lambda(t-1)}||B||_1-\bar{N}+M_t+e_t)  ]_b^c
\end{eqnarray} 
\normalsize

Here $M_t:=||B(t-1)||_1 - \mathbb{E}_{\pi^{(t-1)}} ||B(t-1)||_1$ is a Martingale difference noise sequence, and 
$\lim_{t \rightarrow \infty} e_t=0$. It is easy to see that the derivative of $\mathbb{E}_{\pi_{\beta} | \lambda }||B||_1$ w.r.t. $\lambda$ is bouned for $\lambda \in [b,c]$; hence, $\mathbb{E}_{\pi_{\beta} | \lambda }||B||_1$ is a Lipschitz continuous function of $\lambda$. It is also easy to see that the sequence $\{M_t\}_{t \geq 0}$ is bounded. Hence, by the theory presented in 
\cite[Chapter~$2$]{borkar08stochastic-approximation-book} and 
\cite[Chapter~$5$, Section~$5.4$]{borkar08stochastic-approximation-book}, $\lambda(t)$ converges to the unique zero of $\mathbb{E}_{\pi_{\beta} | \lambda }||B||_1-\bar{N}$ almost surely. Hence,  $\lambda(t) \rightarrow \lambda^*$ almost surely. Since $\lim_{t \rightarrow \infty} d_V(\pi^{(t-1)}, \pi_{\beta| \lambda(t-1)})=0$ and $\pi_{\beta | \lambda}$ is continuous in $\lambda$, the limiting distribution of $B(t)$ becomes $\pi_{\beta | \lambda^*}$.

\section{Proof of Theorem~\ref{theorem:convergence-of-GIBBSLEARNINGEM}}
\label{appendix:proof-of-GIBBSLEARNINGEM}
The proof involves several steps, and the brief outline of these steps are provided one by one. 
\subsubsection{Convergence in the fastest timescale}
Let us denote the probability distribution of $B(t)$ under Algorithm~\ref{algorithm:gibbs-learning-algorithm-for-constrained-problem-EM} by $\pi^{(t)}$ (a column vector indexed by the cofigurations from $\mathcal{B}$), and the corresponding transition probability matrix (TPM) by $A(t)$; i.e., $(\pi^{(t+1)})^T=(\pi^{(t)})^T A(t)=(1-1)\times(\pi^{(t)})^T + 1\times  (\pi^{(t)})^T A(t) $. This form is similar to a standard stochastic approximation scheme as in \cite[Chapter~$2$]{borkar08stochastic-approximation-book} except that the step size sequence for $\pi^{(t)}$ iteration is a constant sequence. Also, if $f^{(t)}(B)$, $\lambda(t)$ and $\theta(t)$ are constant with time $t$, then $A(t)=A$ will also be constant with time $t$, and the stationary distribution for the TPM $A$ will exist and will  be Lipschitz continuous in all (constant) slower timescale iterates. Hence, by using similar argument as in \cite[Chapter~$6$, Lemma~$1$]{borkar08stochastic-approximation-book}, one can show the following for all $B \in \mathcal{B}$:
\begin{equation}\label{eqn:convergence-fastest-timescale}
\lim_{t \rightarrow \infty} |\pi^{(t)}(B)-\pi_{\beta,f^{(t)},\lambda(t), \theta(t)}(B)|=0 \,\,\, a.s.
\end{equation}

\subsubsection{Convergence of iteration \eqref{eqn:fB-update-iid-data}}
Note that,  \eqref{eqn:fB-update-iid-data} depends on $\theta(t)$ and not  on $B(t)$ and $\lambda(t)$; the iteration \eqref{eqn:fB-update-iid-data} depends on $\theta(t)$ through the estimation function $\mu_2(\cdot; \cdot; \cdot)$. Now, $f^{(t)}(B)$ is updated at a faster timescale compared to  $\theta(t)$. Let us consider the iterations \eqref{eqn:fB-update-iid-data} and \eqref{eqn:theta-update-iid-data}; they constitute a two-timescale stochastic approximation. 

Note that, for a given $\theta$, the iteration \eqref{eqn:fB-update-iid-data}  remains bounded inside a compact set independent of $\theta$; hence, using \cite[Chapter~$2$, Theorem~$2$]{borkar08stochastic-approximation-book} with additional modification as suggested in \cite[Chapter~$5$, Section~$5.4$]{borkar08stochastic-approximation-book} for projected stochastic  approximation, we can claim that $\lim_{t \rightarrow \infty} f^{(t)}(B) \rightarrow f_{\theta}(B)$ almost surely for all $B \in \mathcal{B}$, if $\theta(t)$ is kept fixed at a value $\theta$. Also, since $\mu_2(\cdot;\cdot;\theta)$ is Lipschitz continuous in $\theta$, we can claim that $f_{\theta}(B)$ is Lipschitz continuous in $\theta$ for all $B \in \mathcal{B}$. We also have $\lim_{t \rightarrow \infty}\frac{c(\nu(t))}{a(\nu(t))}=0$.

 Hence, by using an analysis similar to that in 
\cite[Appendix~E, Section~C.2]{chattopadhyay-etal15measurement-based-impromptu-deployment-arxiv-v1} (which uses \cite[Chapter~$6$, Lemma~$1$]{borkar08stochastic-approximation-book}), one can claim that:
\begin{equation}\label{eqn:convergence_fB-iid-data}
\lim_{t \rightarrow \infty} |f^{(t)}(B)-f_{\theta(t)}(B)|=0 \,\,\, a.s. \,\,\, \forall B \in \mathcal{B}
\end{equation}
{\em This proves the desired convergence of the iteration \eqref{eqn:fB-update-iid-data}.}

\subsubsection{Convergence of $\lambda(t)$ iteration}
 The $\lambda(t)$ iteration will view $\theta(t)$ as quasi-static and $B(t)$, $f^{(t)}(\cdot)$ iterations as equilibriated.

Let us  assume that $\theta(t)$ is kept fixed at $\theta$. Then, by \eqref{eqn:convergence-fastest-timescale} and \eqref{eqn:convergence_fB-iid-data}, we can work with $\pi_{\beta,f_{\theta},\lambda^{(t)}, \theta}$ in this timescale. 
Under this situation, \eqref{eqn:lambda-update-iid-data} asymptotically tracks the iteration  
$\lambda(t+1)=[\lambda(t)+b(t) ( \sum_{B \in \mathcal{B}} \pi_{\beta,f_{\theta},\lambda(t), \theta}(B) ||B||_1-\bar{N}+M_t )]_0^A$ where $\{M_t\}_{t \geq 0}$ is a Martingale differenece sequence. Now, $\pi_{\beta,f_{\theta},\lambda(t), \theta}(B)$ is Lipschitz continuous in $\theta$ and $\lambda(t)$ (using  Assumption~\ref{assumption:Lipschitz-continuity-wrt-theta}, Assumption~\ref{assumption:existence-of-lambda*} and a little algebra on the expression~\eqref{eqn:definition-of-Gibbs-distribution}). If $A_0$ is large enough, then, by the theory of \cite[Chapter~$2$, Theorem~$2$]{borkar08stochastic-approximation-book} and \cite[Chapter~$5$, Section~$5.4$]{borkar08stochastic-approximation-book}, one can claim that $\lambda(t) \rightarrow \lambda^*(\theta)$ almost surely, and $\lambda^*(\theta)$ is Lipschitz continuous in $\theta$ (by Assumption~\ref{assumption:existence-of-lambda*}).

 Hence, by using similar analysis as in 
\cite[Appendix~E, Section~C.2]{chattopadhyay-etal15measurement-based-impromptu-deployment-arxiv-v1} (which uses  \cite[Chapter~$6$, Lemma~$1$]{borkar08stochastic-approximation-book}), we can say that,  under iteration~\eqref{eqn:lambda-update-iid-data}:
\begin{equation}\label{eqn:convergence-in-lambda-iteration-iid-data}
\lim_{t \rightarrow \infty} |\lambda(t)-\lambda^*(\theta(t))|=0 \,\,\, a.s.
\end{equation}

\subsubsection{Convergence of the $\theta(t)$ iteration}
Note that, \eqref{eqn:theta-update-iid-data} is the slowest timescale iteration and hence it will view all other there iterations (at three different timescales) as equilibriated. However, this iteration is not affected by other iterations. Hence, this iteration is an example of simultaneous perturbation stochastic approximation as in \cite{spall92original-SPSA} (since $\theta^*$ lies inside the interior of $\Theta$), but with a projection operation applied on the iterates.  Hence, by combining \cite[Proposition~$1$]{spall92original-SPSA} and the discussion in \cite[Chapter~$5$, Section~$5.4$]{borkar08stochastic-approximation-book}, we can say that $\lim_{t \rightarrow \infty} \theta(t)=\theta^*$ almost surely.

\subsubsection{Completing the proof}
We have seen that  $\lim_{t \rightarrow \infty} \theta(t)=\theta^*$ almost surely. Hence, by \eqref{eqn:convergence-in-lambda-iteration-iid-data}, $\lim_{t \rightarrow \infty} \lambda(t)=\lambda^*(\theta^*)$ almost surely. By \eqref{eqn:convergence_fB-iid-data}, $\lim_{t \rightarrow \infty} f^{(t)}(B)=f_{\theta^*}(B)$ almost surely for all $B \in \mathcal{B}$. Then, by \eqref{eqn:convergence-fastest-timescale}, $\lim_{t \rightarrow \infty} \pi^{(t)}(B)=\pi_{\beta,f_{\theta^*},\lambda^*(\theta^*), \theta^*}(B)$ almost surely. Hence, Theorem~\ref{theorem:convergence-of-GIBBSLEARNINGEM} is proved.

\end{document}